\documentclass[aps,prd,superscriptaddress,floatfix,nofootinbib,notitlepage]{revtex4-1}

\usepackage{graphicx}
\usepackage{amsmath}
\usepackage{longtable}
\usepackage{aas_macros}
\usepackage{xcolor}
\usepackage{xspace}

\graphicspath{{./figures/}}

\usepackage{lineno}

\usepackage[sort&compress]{natbib}
\usepackage{hyperref}
\newcommand{\gammaRay}{$\gamma$ ray}
\newcommand{\gammaRays}{$\gamma$ rays}
\newcommand{\gammaRayHyph}{$\gamma$-ray}

\newcommand{\stools}{\emph{ScienceTools}}

\newcommand{\selection}[1]{\texttt{#1}}

\newcommand{\irf}[1]{\texttt{#1}}

\newcommand{\fov}{{\rm Fo\kern-1ptV}}

\newcommand{\probE}{\ensuremath{P_{\rm E}}}

\newcommand{\Secref}[1]{Section~\ref{#1}}
\newcommand{\secref}[1]{Sec.~\ref{#1}}

\newcommand{\appref}[1]{App.~\ref{#1}}

\newcommand{\tabref}[1]{Tab.~\ref{tab:#1}}
\newcommand{\Figref}[1]{Figure~\ref{fig:#1}}
\newcommand{\figref}[1]{Fig.~\ref{fig:#1}}

\newcommand{\Eqref}[1]{Eqn.~\eqref{#1}}

\newcommand{\Fermi}{{\textit{Fermi}}}
\newcommand{\fermi}{\Fermi}

\newcommand{\peight}{\irf{Pass~8}}

\newcommand{\Etrue}{\ensuremath{E}}
\newcommand{\Egamma}{\ensuremath{E_{\gamma}}}
\newcommand{\Ereco}{\ensuremath{E'}}
\newcommand{\Epivot}{\ensuremath{E_{0}}}

\newcommand{\AnnGG}{\ensuremath{\langle\sigma v\rangle_{\gamma\gamma}}}
\newcommand{\Ann}{\ensuremath{\langle\sigma v\rangle}}

\newcommand{\newText}[1]{{\color{black}{#1}}}

\newcommand{\eg}{e.g.\xspace}

\newcommand{\beff}{\ensuremath{b_{\text{eff}}}}
\newcommand{\fstat}{\ensuremath{\delta f_{\text{stat}}}}
\newcommand{\fsyst}{\ensuremath{\delta f_{\text{syst}}}}

\newcommand{\intROI}{\ensuremath{\int^{ROI}}}

\newcommand{\GamBkg}{\ensuremath{\Gamma_{\rm bkg}}}

\newcommand{\nSig}{\ensuremath{n_{\rm sig}}}
\newcommand{\nSigBF}{\ensuremath{n_{\rm sig}'}}
\newcommand{\nSys}{\ensuremath{n_{\rm syst}}}
\newcommand{\nBkg}{\ensuremath{n_{\rm bkg}}}

\newcommand{\PhiMono}{\ensuremath{\Phi_{\text{mono}}}}

\newcommand{\sigmaSys}{\ensuremath{\sigma_{\rm syst}}}

\newcommand{\CSig}{\ensuremath{C_{\rm sig}}}
\newcommand{\CBkg}{\ensuremath{C_{\rm bkg}}}

\newcommand{\slocal}{\ensuremath{s_{\rm local}}}
\newcommand{\sglobal}{\ensuremath{s_{\rm global}}}

\newcommand{\smax}{\ensuremath{s_{\rm max}}}

\newcommand{\Deff}{\ensuremath{D_{\rm eff}}}

\newcommand{\Exposure}{\ensuremath{\mathcal{E}}}

\newcommand{\Rgc}{\ensuremath{R_{\rm GC}}}

\newcommand{\sigE}{\ensuremath{\sigma_{E}}}

\ifdefined\bwfigures
\newcommand{\colSw}[1]{#1_bw} 
\else
\newcommand{\colSw}[1]{#1} 
\fi

\newlength{\enumindent}
\newcounter{fermicnt}
% ...

\newlength{\twocolfigwidth}
\newlength{\onecolfigwidth}
\newlength{\twothirdscolfigwidth}
\newcommand{\onewidepanel}[3]{%
  \begin{figure}[#1]
    \centering
    \includegraphics[width=\twocolfigwidth,clip=true,trim=0 5 0 5]{#2}
    
    #3
  \end{figure}
}

\newcommand{\onepanel}[3]{%
  \begin{figure}[#1]
    \centering
    \includegraphics[width=\onecolfigwidth]{#2}
    
    #3
  \end{figure}
}

\newcommand{\twopanel}[4]{%
  \begin{figure}[#1]
    \centering
    \includegraphics[width=\onecolfigwidth]{#2}\hfill%
    \includegraphics[width=\onecolfigwidth]{#3} 

    #4
  \end{figure}
}

\newcommand{\fourpanel}[6]{%
  \begin{figure}[#1]
    \centering
    \includegraphics[width=\onecolfigwidth]{#2}\hfill%
    \includegraphics[width=\onecolfigwidth]{#3}\\
    \includegraphics[width=\onecolfigwidth]{#4}\hfill%
    \includegraphics[width=\onecolfigwidth]{#5}

    #6
  \end{figure}
}

\begin{document}

\title{Updated Search for Spectral Lines from Galactic Dark Matter Interactions with Pass 8 Data from the Fermi Large Area Telescope}

\date{\today}

\author{M.~Ackermann}
\affiliation{Deutsches Elektronen Synchrotron DESY, D-15738 Zeuthen, Germany}
\author{M.~Ajello}
\affiliation{Department of Physics and Astronomy, Clemson University, Kinard Lab of Physics, Clemson, SC 29634-0978, USA}
\author{A.~Albert}
\email{aalbert@slac.stanford.edu}
\affiliation{W. W. Hansen Experimental Physics Laboratory, Kavli Institute for Particle Astrophysics and Cosmology, Department of Physics and SLAC National Accelerator Laboratory, Stanford University, Stanford, CA 94305, USA}
\author{B.~Anderson}
\affiliation{Royal Swedish Academy of Sciences Research Fellow, funded by a grant from the K. A. Wallenberg Foundation}
\author{W.~B.~Atwood}
\affiliation{Santa Cruz Institute for Particle Physics, Department of Physics and Department of Astronomy and Astrophysics, University of California at Santa Cruz, Santa Cruz, CA 95064, USA}
\author{L.~Baldini}
\affiliation{Universit\`a di Pisa and Istituto Nazionale di Fisica Nucleare, Sezione di Pisa I-56127 Pisa, Italy}
\affiliation{W. W. Hansen Experimental Physics Laboratory, Kavli Institute for Particle Astrophysics and Cosmology, Department of Physics and SLAC National Accelerator Laboratory, Stanford University, Stanford, CA 94305, USA}
\author{G.~Barbiellini}
\affiliation{Istituto Nazionale di Fisica Nucleare, Sezione di Trieste, I-34127 Trieste, Italy}
\affiliation{Dipartimento di Fisica, Universit\`a di Trieste, I-34127 Trieste, Italy}
\author{D.~Bastieri}
\affiliation{Istituto Nazionale di Fisica Nucleare, Sezione di Padova, I-35131 Padova, Italy}
\affiliation{Dipartimento di Fisica e Astronomia ``G. Galilei'', Universit\`a di Padova, I-35131 Padova, Italy}
\author{R.~Bellazzini}
\affiliation{Istituto Nazionale di Fisica Nucleare, Sezione di Pisa, I-56127 Pisa, Italy}
\author{E.~Bissaldi}
\affiliation{Istituto Nazionale di Fisica Nucleare, Sezione di Bari, 70126 Bari, Italy}
% ...
\begin{abstract}
This list is preliminary; the status is not yet "ready to submit"
\end{abstract}

\begin{abstract}
\pagebreak
Dark matter in the Milky Way may annihilate directly into \gammaRays, producing a monoenergetic spectral line.
Therefore, detecting such a signature would be strong evidence for dark matter annihilation or decay. We search for spectral lines in the $Fermi$ Large Area Telescope 
observations of the Milky Way halo in the energy range 200 MeV -- 500 GeV using analysis methods from our most recent line searches. 
The main improvements relative to previous works are our use of 5.8 years of data reprocessed with the \irf{Pass 8} event-level analysis 
and the additional data resulting from the modified observing strategy designed to increase exposure of the Galactic center region. We searched in five sky 
regions selected to optimize sensitivity to different theoretically-motivated dark matter scenarios and find no significant detections. In addition to presenting the results from our search for lines, we also investigate the previously reported tentative detection of a line at 133 GeV using the new \irf{Pass 8} data. 
  
\end{abstract}
\pacs{95.35.+d,95.30.Cq,98.35.Gi}
\maketitle

\section{INTRODUCTION}\label{sec:intro}

\newText{Cosmological observations reveal} that $\sim$80\% of the matter in the Universe is dark matter (DM). 
Cosmic microwave background measurements, galactic rotation curves, gravitational lensing (among others) each provide strong 
evidence for the existence of DM~\cite{Ade:2013lta,Sofue:2000jx,Clowe:2006eq}. 
One of the leading DM candidates are the Weakly Interacting Massive Particles (WIMP), though other candidates such as 
gravitinos may also account for some, or all of the observed DM (see Ref.~\cite{Bertone:2004pz,Feng:2010gw} for recent reviews of particle DM candidates).

WIMPs and other DM candidates may annihilate or decay to Standard Model particles, which would produce \gammaRays. Gamma-ray signals from 
DM annihilation or decay are expected to typically produce a broad spectral signature (see Ref.~\cite{Bringmann:2012ez,Ibarra:2013cra} for recent 
indirect DM search reviews). The difficulty in detecting such a signal lies in distinguishing it
from other standard astrophysical processes. 
However, if the DM annihilates or decays into a photon and a neutral particle (such as another photon or Z boson), approximately monoenergetic \gammaRays\
will be produced in the rest frame. For non-relativistic DM particles, this would give rise to a monoenergetic photon signal in the otherwise smooth
spectrum of the standard astrophysical emission. 
Such a sharp spectral signature is not expected from standard astrophysical mechanisms, \newText{though non-DM induced mechanisms have been proposed~\citep[\newText{\eg,}][]{Aharonian:2012}}.  
The branching fraction of monoenergetic DM annihilation channels is typically loop-suppressed, $\Ann_{\gamma\gamma}\sim10^{-4}\Ann-10^{-1}\Ann $, where \Ann\ is 
the total velocity-averaged DM annihilation cross section and $\Ann_{\gamma\gamma}$ is the cross section for DM annihilation to two 
\gammaRays~\cite{REF:Bergstrom:1997fh,REF:Matsumoto:2005ui,REF:Ferrer:2006hy,REF:Gustafsson:2007pc,REF:Profumo:2008yg}. The total annihilation cross section that would produce the currently
observed abundance of DM in the Universe is $\Ann\approx3\times10^{-26}~\rm{cm}^{-3}\rm{s}^{-1}$, assuming that DM is a thermal relic~\cite{Steigman:2012nb}.

In this paper, we search for \gammaRayHyph\ spectral lines using data obtained by the Large Area Telescope (LAT)~\cite{REF:2009.LATPaper} on board the \Fermi~\gammaRayHyph~Space Telescope.  
The LAT has been surveying the \gammaRayHyph\ sky in an energy range of 20 MeV to over 300 GeV since 2008, \newText{making it an ideal instrument to search for \gammaRays\ from DM interactions. WIMP DM candidates typically have $\sim$GeV to $\sim$TeV scale masses; therefore, WIMP annihilations would produce \gammaRays\ detectable by the LAT. Specifically, DM annihilations directly into pairs of \gammaRays\ will create a spectral line at the rest mass energy of the DM particle. Besides being a strong indication of WIMP interactions, spectral lines are one of the best ways to search for gravitino (another DM candidate) decay~\cite{REF:Ibarra:2007wg}.} 

The LAT Collaboration has published four previous line searches, each improving and expanding the analysis relative to the previous works. 
Our first analysis searched for lines from 30 to 200 GeV using 11 months of data~\cite{REF:2010.LineSearch}. The search using two years of data expanded the energy range to search for lines down to 
7 GeV and made use of control regions to estimate systematic uncertainties~\cite{REF:2012.LineSearch}.  
The 3.7 year analysis covered the energy range from 5 to 300 GeV and included a thorough investigation and quantification of the systematic uncertainties~\cite{REF:P7Line}.  
That work also included a detailed investigation of a tentative line-like feature near 133 GeV reported in the region around the Galactic center (GC)~\cite{Bringmann:2012vr,REF:Weniger:2012tx}. 
An analysis extending to lower energies (100 MeV to 10 GeV), where systematic uncertainties typically dominate, was performed using 5.2 years of data~\cite{REF:LELine}. In that paper, the systematic 
uncertainties were quantified, and also incorporated in the likelihood analysis.
These past analyses used previous iterations (or `Passes') of the LAT
event reconstruction and classification~\cite{REF:2009.LATPaper, REF:2012.P7Perf}. 

In this work, we update the results from our previous line searches using 5.8 years of \peight~data.  We present an analysis performed over three decades in energy that incorporates systematic uncertainties estimated using control regions. 
\Secref{sec:method_data}, \secref{sec:method_roi}, and \secref{sec:method_edisp} outline the event selection, regions of interest, and energy dispersion modeling respectively.  
\Secref{sec:fitting} describes the methods for fitting the \gammaRayHyph\ spectrum and \secref{sec:syst} details the systematic studies. 
\Secref{sec:results} presents the results of searching for a monoenergetic line in the \gammaRayHyph\ spectrum. 
\Secref{sec:133GeV_Feature} goes into greater detail on the previously reported tentative detection near 133 GeV in the region around the GC. Finally, \secref{sec:summary} summarizes the results.

\section{DATA AND EVENT SELECTION}\label{sec:method_data}

The LAT Collaboration has recently developed an extensive update of the event reconstruction and classification (\peight) 
which improves the performance of the LAT and reduces systematic uncertainties~\cite{Atwood:2013rka}. 
\peight\ events are classified according to the estimated accuracy of the direction reconstruction.  
Each event was assigned a PSF `type' from PSF0 (worst) to PSF3 (best), with each type containing about one quarter of the total 
number of events in logarithmic energy bins. Events are also classified according to the estimated accuracy of the energy reconstruction.  
Each event is assigned an energy dispersion `type' ($e_{t}$) from EDISP0 (worst) to EDISP3 (best) according to the reconstruction quality estimator `BestEnergyProb'.  
Each energy dispersion event type contains roughly one quarter of the events in log(E/MeV) energy bins. 
\Figref{EDispType} (left) shows \newText{the energy dispersion} \Deff\ ($(\Ereco - \Etrue) / \Etrue$) at 100 GeV for each EDISP type.  The energy resolution (68\% containment) is shown in \figref{EDispType}~(right).  
The expected energy resolution is greatly improved in EDISP3 compared to EDISP0.  

\twopanel{ht}{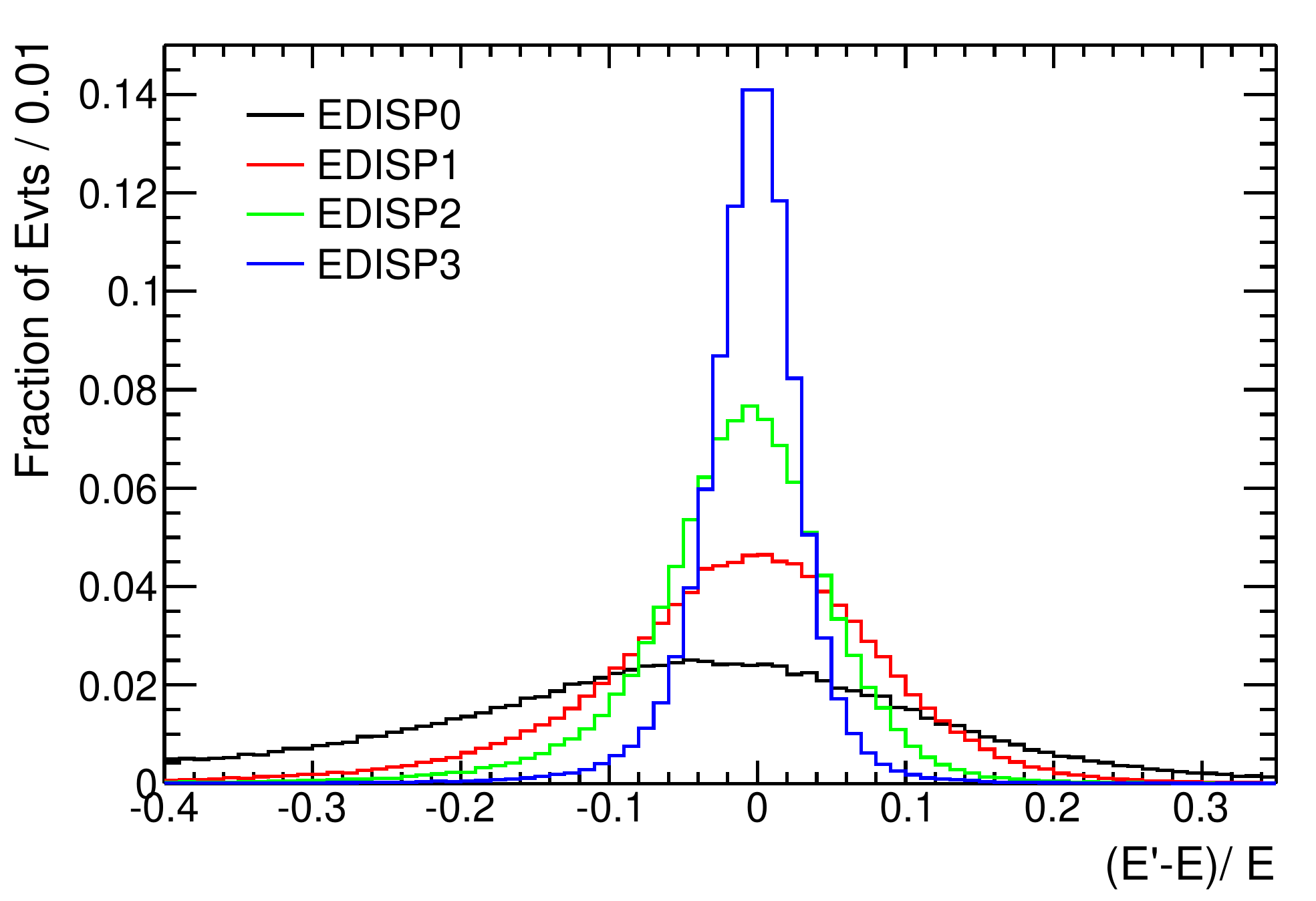}{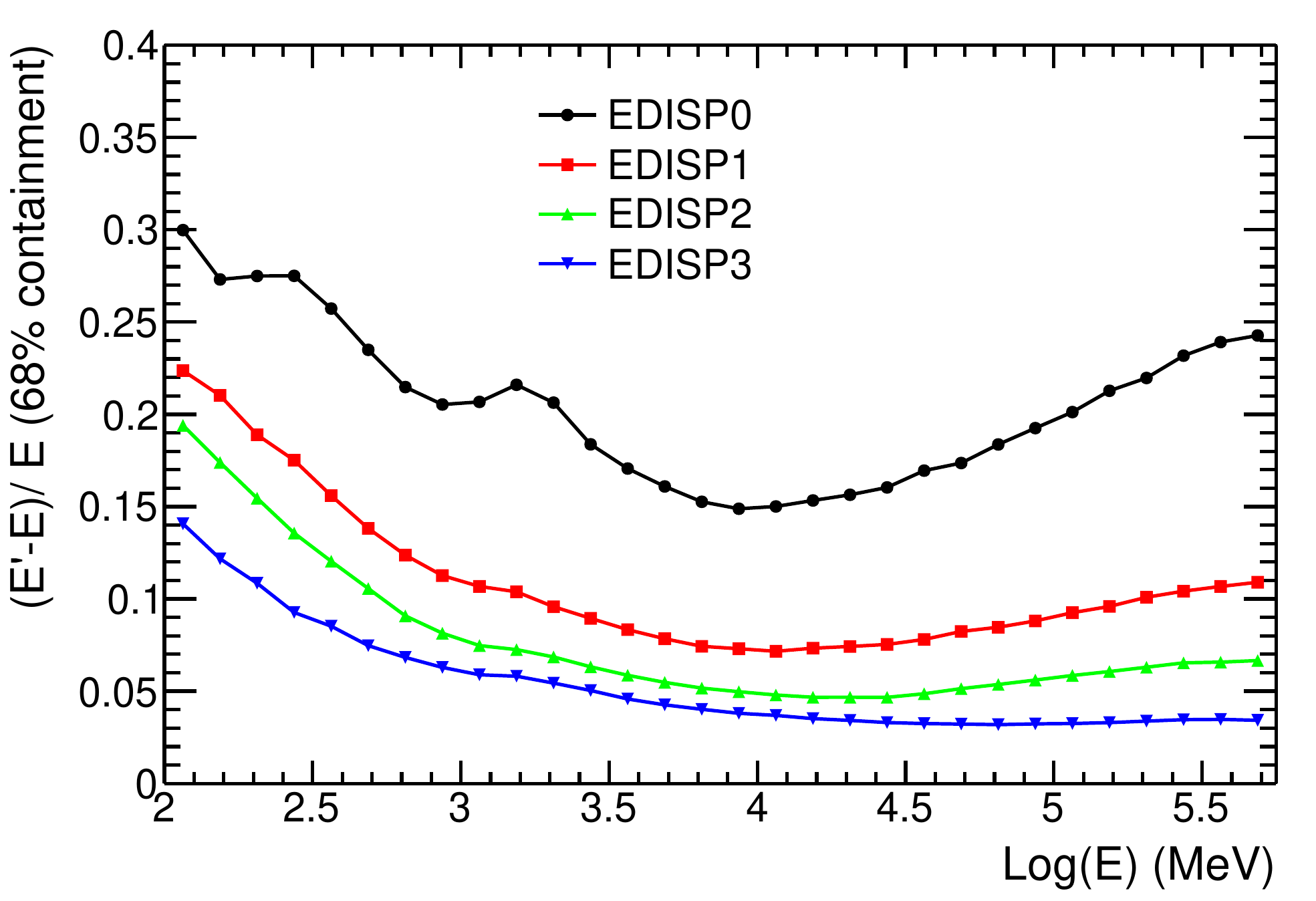}
{\caption{\label{fig:EDispType}
\irf{P8\_CLEAN\_V5} energy dispersion at 100 GeV (left) and energy resolution as a function of energy (right) for the four event types. \Etrue\ is the true energy of the generated~\gammaRay\ and~\Ereco\ is the reconstructed energy.}}

\Figref{EDispP7P8100GeV} compares the energy dispersion 
in \irf{Pass 8} and \irf{Pass 7REP} for the energy range $5<\rm{log(E)}/\rm{MeV}<5.2$. While the energy resolution has not significantly improved in \irf{Pass 8}, 
the number of events accepted in each event class is higher.  The increase in acceptance is due, in part, to the ability of event reconstruction in 
\irf{Pass 8} to isolate the gamma-ray events when a cosmic ray is in near coincidence with the gamma ray, 
more generally referred to as pile-up or `ghost' events~\cite{REF:2012.P7Perf}. 
The largest impact relevant to this analysis is an increased effective area in \irf{Pass 8} by $\sim30$\% for events above 10 GeV for the Clean class. 
We note that the specific `Clean' event selections are different between \irf{Pass 7REP} and \irf{Pass 8}. 
However, both are defined such that the resulting residual CR rate is at or below the extragalactic \gammaRayHyph\ background above 100 MeV.  

\onepanel{ht}{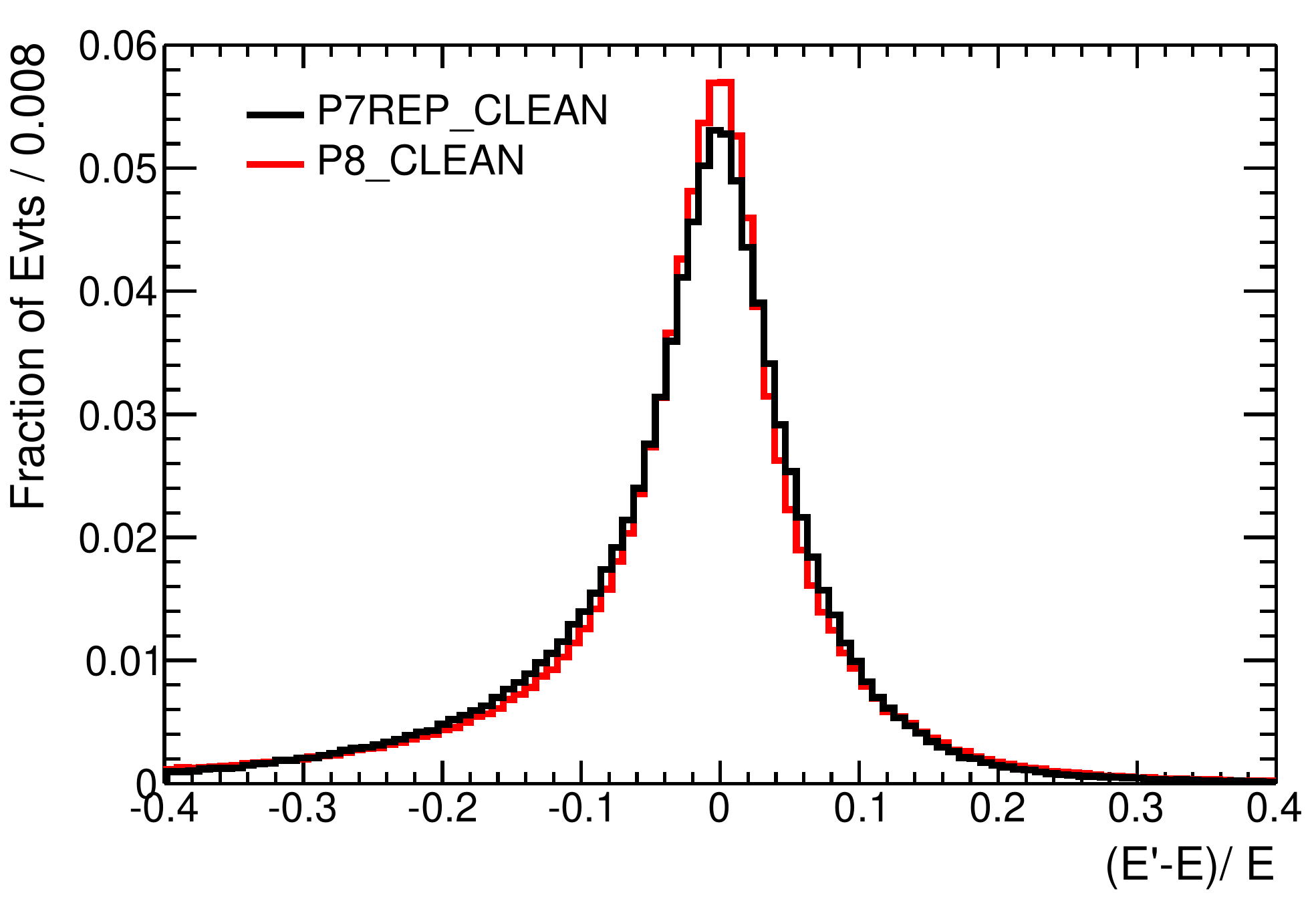}
{\caption{\label{fig:EDispP7P8100GeV} Normalized energy dispersion in the range $5<\rm{log(E)}/\rm{MeV}<5.2$ from MC with \irf{CLEAN} event class selection for \irf{Pass 8} and \irf{Pass 7REP} respectively.  \Etrue\ is the true energy of the generated~\gammaRay\ and~\Ereco\ is the reconstructed energy.}}

We used data from a 5.8 year period with the \irf{P8\_CLEAN} event selection to search for spectral lines in the energy range 200 MeV to 500 GeV. 
To allow for sidebands in each fit, we used LAT \gammaRayHyph~data from 100 MeV to 750 GeV. 
Details on additional event selections are given in \tabref{selection}.
We split the data into Celestial and Earth Limb datasets using cuts on both the instrument rocking angle ($\theta_r$) and the event zenith angle ($\theta_z$); see Fig. 1 of Ref.~\cite{REF:P7Line} for a schematic depicting $\theta_r$, $\theta_z$, and event incident angle $\theta$.  
Gamma rays from the Earth's Limb are produced by cosmic-ray (CR) interactions in the Earth's upper atmosphere so it is used as a control region where no true spectral lines are expected from 
dark matter annihilation or decay (see \appref{app:control_samples}). 

We extract our signal regions of interest (ROIs) and other control regions from the Celestial data set, where we require $\theta_z < 100^{\circ}$ in order to remove emission from the bright Earth's Limb.
No point source masking is applied since at lower energies the width of the point spread function (PSF) increases significantly.  Using an energy-dependent point source mask, as was done in our 3.7-year analysis, would result in removing most of the dataset below $\sim$1 GeV. 
The initial data reduction and all of the exposure calculations were performed using the LAT \stools\footnote{The \stools\
and documentation are available at \url{http://fermi.gsfc.nasa.gov/ssc/data/analysis/scitools/overview.html}} version 09-33-03, and the \irf{P8\_CLEAN\_V5}\ instrument response 
functions\footnote{The IRF used (\irf{P8\_CLEAN\_V5}) is not one that will be released publicly. \irf{P8\_CLEAN\_V6}\ will be the first released. Both have been derived with 
the same data set and have only minor technical differences that do not affect this analysis.}.

\begin{table}[ht]
  \caption{\label{tab:event_samples}Summary table of data selections.}
  \begin{center}
  \begin{tabular}{lcc}
    \hline\hline
\label{tab:selection}
Selection & Celestial & Earth Limb \\
\hline
Observation period & August 4, 2008 -- April 30, 2014 & August 4, 2008 -- April 30, 2014 \\
Mission Elapsed Time\footnote{Mission Elapsed Time is the number of seconds since 00:00:00 UTC January 1, 2000.} (s) & [239557417--420595073] & [239557417--420595073] \\
Energy Range (GeV) & [0.1--750] & [2.5--750]  \\
Rocking Angle ($\theta_r$) & $<$ 52$^{\circ}$ & $>$ 52$^{\circ}$   \\
Zenith Angle ($\theta_z$) & $< 100^{\circ}$ & $111^{\circ} < \theta_z < 113^{\circ}$\\
Data quality\footnote{\texorpdfstring{\selection{DATA\_QUAL == 1 \&\& LAT\_CONFIG ==1}}{DATA_QUAL == 1 \&\& LAT_CONFIG == 1}} & Yes & Yes  \\
\hline\hline
\end{tabular}
\end{center}
\end{table}

\section{REGIONS OF INTEREST}\label{sec:method_roi}  
Within our Celestial dataset, we define our signal ROIs to be the same as those considered in Ref.~\cite{REF:P7Line}. 
These ROIs were optimized for either annihilating or decaying DM and for different profiles of the spacial distribution of the DM distribution in the Galaxy.
We used four models of the DM distribution: Navarro-Frenk-White (NFW)~\cite{REF:1996ApJ...462..563N}, an adiabatically contracted NFW (NFWc), 
Einasto~\cite{REF:2010MNRAS.402...21N}, and a cored isothermal profile~\cite{REF:1980ApJS...44...73B}. 

A generalized NFW profile~\cite{Kravtsov:1997dp} is given by

\begin{equation}
  \rho(r) = \frac{\rho_0}{(r/r_s)^{\gamma}(1+r/r_s)^{3-\gamma}}
\end{equation}

\noindent with $r_s = 20$ kpc. The NFW and NFWc correspond, respectively, to the cases where $\gamma = 1$ and $\gamma = 1.3$. We also use an Einasto profile defined as:

\begin{equation}
  \rho(r) = \rho_0 \exp\{ -(2/\alpha)[(r/r_s)^\alpha - 1]\},
\end{equation}

\noindent where $r_s=20$ kpc and $\alpha=0.17$~\cite{REF:2010MNRAS.402...21N}.
Finally, we consider a cored, isothermal profile given by: 

\begin{equation}
  \rho(r) = \frac{\rho_0}{1+(r/r_s)^2}
\end{equation}

\noindent with $r_s=5$ kpc.
We normalize all profiles by fixing the local DM density to $\rho(r_\odot=8.5$~kpc$)$ = 0.4~GeV~cm$^{-3}$~\footnote{We note that values ranging from 0.2 -- 0.85~GeV~cm$^{-3}$ are possible at the present~\cite{REF:2010JCAP...08..004C,Salucci:2010qr,2012MNRAS.425.1445G}. Assuming a different local DM density would simply scale our limits for DM annihilation and decay by a factor inversely proportional to $\rho(r_\odot)^2$ or $\rho(r_\odot)$ respectively}. \Figref{DMprofile} compares these DM distributions. 

\onepanel{ht}{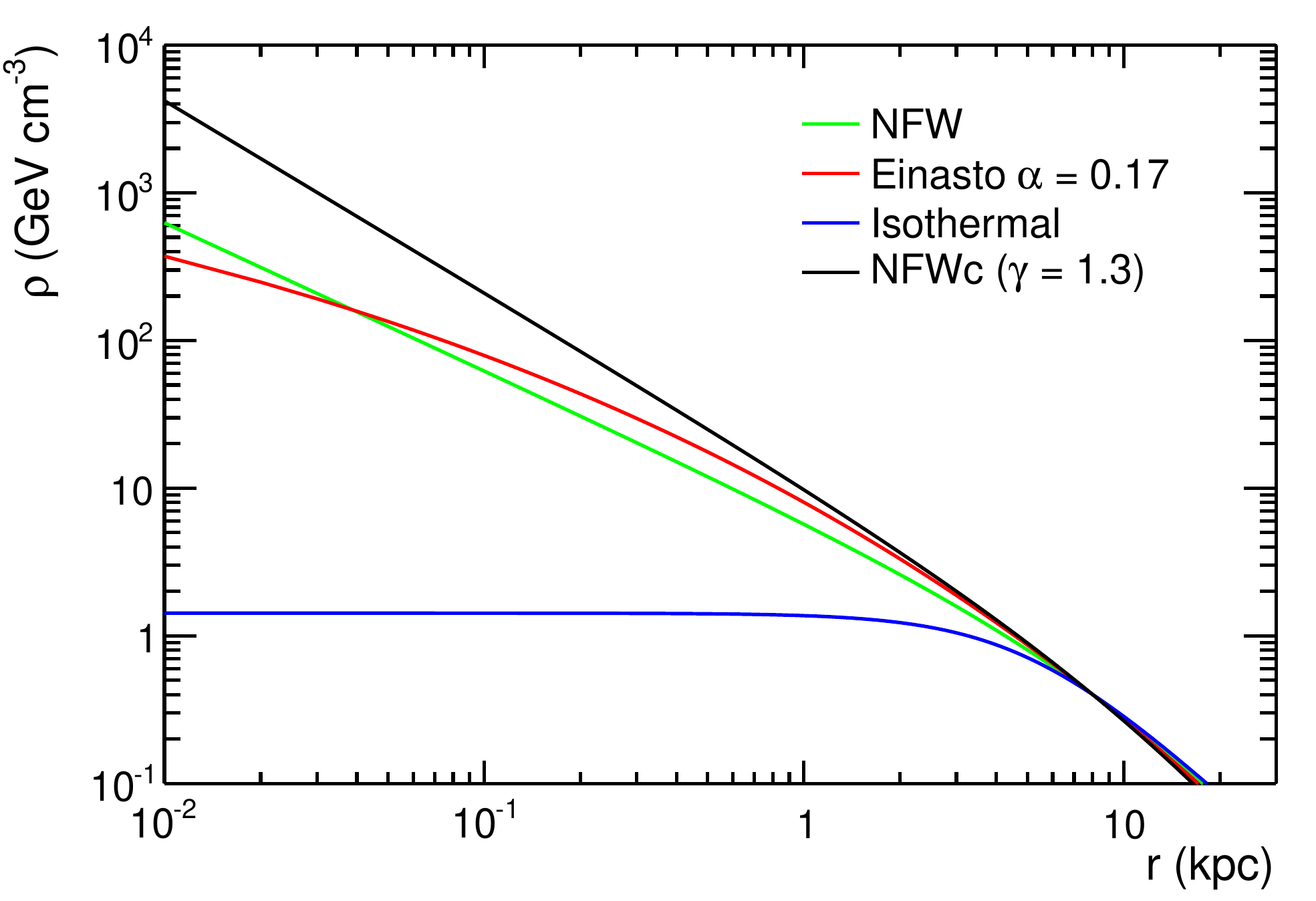}{
\caption{\label{fig:DMprofile} The four DM profiles as a function of the distance from the GC, $r$.}}

Each ROI is defined as a circular region centered on the GC. We mask the Galactic plane (GP) except for a $12^{\circ}\times10^{\circ}$ box \newText{centered on the GC}. 
The ROIs are named after the value of \Rgc\ they subtend (e.g., for R3, $\Rgc=3^{\circ}$).
The annihilation ROIs are R3, R16, R41, and R90 optimized for the NFWc, Einasto, NFW, and Isothermal DM profiles respectively. 
R180 is \newText{the optimal ROI} to search for spectral lines from DM decay (e.g. $\chi\rightarrow\nu\gamma$). 
In the GP, longitudes further than $6^{\circ}$ from the GC are removed from all ROIs larger than R3. 
This is because we do not expect a large DM signal in that region and the \gammaRayHyph\ emission is dominated by standard astrophysical sources.
\newText{For details on the ROI-optimization procedure, see App. B of ref~\cite{REF:P7Line}}.
 
\Figref{ROIs} shows the counts map of the Celestial dataset in R180 with the outlines of the other signal ROIs, and the exclusion of the GP.
We note that these ROIs are different from those in Ref.~\cite{REF:LELine}, which optimized the ROIs only at lower energies where the fits are dominated by systematic uncertainties. 

\onewidepanel{ht}{\colSw{Figure_4}}{
\caption{\label{fig:ROIs} Counts map for the Celestial dataset
  binned in $1^{\circ} \times 1^{\circ}$ spatial bins in the R180
  ROI, and plotted in Galactic coordinates using the Hammer-Aitoff
  projection.  The energy range is 1--750~GeV.  Also shown are the outlines of
  the other ROIs (R3, R16, R41, and R90) used in this search. The GP region with longitude greater than $6^{\circ}$ from the GC and latitude smaller than $5^{\circ}$ is removed from all signal ROIs. }}

For the smallest ROI, R3, the effects of leakage both in and out of the ROI from the PSF dominate the search region. 
The 68\% containment radius of PSF3 is $\sim2^{\circ}$ at 200 MeV and improves to $\sim0.4^{\circ}$ at 1 GeV. 
Therefore, we only use PSF3 events for fits in R3 below 1 GeV. For all other fits, events from every PSF type are used.

\section{ENERGY DISPERSION MODELING}\label{sec:method_edisp}
The energy dispersion is the probability density of reconstructing a true energy (\Etrue) as \Ereco. 
The LAT Collaboration parametrizes the energy dispersion probability density function of \Ereco\ in cos$\theta$ and true energy. In this work, we define an effective energy dispersion (\Deff), where we have averaged over cos$\theta$ assuming an isotropic 
source\footnote{Our effective energy dispersion \Deff\ can be evaluated using the publicly available LAT energy dispersion models by performing an acceptance-weighted averaging along cos$\theta$.}.

As in Ref.~\cite{REF:P7Line}, we find \Deff\ for a given \Etrue\ is well-described as a sum of three Gaussians.  We create separate \Deff~models for each of the EDISP types. Specifically, this is done as follows: 

\begin{equation}\label{eq:TripGaus}
\Deff(\Ereco;\Etrue,e_{t}) = \sum_{k=1}^3
\frac{a_k}{\sigma_k\sqrt{2\pi}}e^{-((\Ereco/\Etrue) - (1+\mu_k))^2/2\sigma_k^2}\;,
\end{equation}

\noindent where $a_3 = 1-a_2-a_1$ (with $a_i > 0$ required) and $\sigma_1 > \sigma_2 > \sigma_3$. 
We fit the triple Gaussian model at energies from 100 MeV $<$ \Etrue\ $<$ 1 TeV in logarithmic steps of 0.25. 
Then we can define \Deff~for any energy by interpolating the parameters of the Gaussian.
This method differs slightly from that in Ref~\cite{REF:P7Line} by using EDISP type as the second variable (in addition to \Etrue) in the ``2D" \Deff~model instead of \probE. 
By modeling the energy dispersion separately for each EDISP type, we are able to give higher weight to events with a better energy reconstruction.
Using the EDISP types adds extra information in the fit and improves the statistical power 
over a ``1D" model by $\sim$10--15\% depending on energy. 

\section{FITTING}\label{sec:fitting}
\subsection{Fitting Procedure}\label{subsec:fitting}
To fit for spectral lines, we use a maximum 
likelihood procedure in sliding energy windows 
in each of the five ROIs described in \secref{sec:method_roi}. We fit at a fixed \Egamma\ at the center of the energy window.  We increment \Egamma\ in steps of 0.5 \sigE(\Egamma), where \sigE(\Egamma) is the energy resolution (68\% containment) of the LAT at \Egamma. 
We perform our fits in the energy domain and
define both a background spectrum model (\CBkg) and a signal spectrum model (\CSig). We do not incorporate spatial information in our fits since it would make the resulting flux limits dependent on the DM distribution profile assumed. Rather, we perform a generic search for monoenergetic signals in each ROI.
Since we fit in narrow energy windows, we approximate the gamma-ray background from diffuse and point sources
as a simple power law. The resulting expected distribution of counts is:

\begin{equation}\label{eq:BkgPDF_1D}
\CBkg(\Ereco|\GamBkg,\nBkg) = \alpha\left(\frac{\Ereco}{\Epivot}\right)^{-\GamBkg}\mathcal{E}(\Ereco),
\end{equation}

\noindent where \GamBkg\ is the power-law index, \Epivot\ is a reference energy set to 100~MeV,
and $\mathcal{E}(\Ereco)$ is the energy-dependent exposure averaged over each ROI, which is 
needed since the fit is performed in count space. The normalization factor $\alpha$ is defined such that the total number of background events in the fit window is $\nBkg = \int \alpha \left(\frac{\Ereco}{\Epivot}\right)^{-\GamBkg}\mathcal{E}(\Ereco) d\Ereco$.  
We did not explicitly convolve our background model with the energy dispersion (i.e. for \CBkg\ we assume \Ereco = \Etrue). 
For fits below 200 MeV, this approximation is not valid and significantly degrades the goodness of the fits. Therefore, we limit our search range to \Egamma\ $>$ 200 MeV.

Our signal spectrum is $\CSig(\Ereco|\Egamma) = \nSigBF \Deff(\Ereco|\Egamma)$. 
We account for systematic uncertainties that may induce a false line-like signal or mask a true line-like signal in our fitting by using the procedure described in Ref.~\cite{REF:LELine}. This is especially important for fits with very small statistical uncertainties (see below).
We include a nuisance parameter by treating the best-fit number of signal events (\nSigBF) as the sum of the true number of signal events (\nSig) and a systematic offset (\nSys) such that \nSig = \nSigBF - \nSys. 
We constrain \nSys\ by modeling it as a Gaussian with a fixed width \sigmaSys\ with zero mean, where~\newText{$\sigmaSys = \fsyst\times\beff$ and} is determined based on fits in control regions (see \secref{sec:syst}). 

Our overall model to fit for a line at \Egamma\ is:

\begin{equation}\label{eq:SigPlusBkgCnts}
C(\Ereco|\vec{\beta}) = 
\left(\nSigBF \Deff(\Ereco|\Egamma)  + \alpha
\left(\frac{\Ereco}{\Epivot}\right)^{-\GamBkg}\mathcal{E}(\Ereco) \right)
\times G_{\rm syst}(\nSys,\beff),
\end{equation}

\newText{

where

\begin{equation}
G_{\rm syst}(\nSys,\beff) = \frac{1}{\sigmaSys\sqrt{2\pi}} e^{-\nSys^{2}/2\sigmaSys^{2}},
\end{equation}
}

\noindent \Deff($\Ereco|\Egamma$) is a weighted sum over the four EDISP types\footnote{We weight using the observed distribution of EDISP types in each fit ROI and energy window.}, and $\vec{\beta}$ represents the model parameters \Egamma, \GamBkg, \nSigBF, and \nBkg. 


Unbinned fits are performed when there are fewer than 10,000 events in the fit window. When there are more than 10,000 events, 
we bin the data in 63 energy bins to avoid large computation times. Each energy bin is much narrower than the energy resolution of the LAT, 
making the binned fits a close approximation to the unbinned fits, which we confirmed in several test cases. 
We fit for a monoenergetic signal, and our results are applicable to any sharp spectral feature much narrower than the LAT energy resolution. 
We will interpret our results in the context of DM annihilation or decay in the next section.

To discuss uncertainties involved in a line search, we define a quantity called the \textit{fractional signal} ($f$)~\cite{REF:LELine,REF:P7Line}, which can be thought of as the fractional size of a line-like signal around the peak.  Specifically $f$ is

\begin{equation}
    f\equiv n_{\text{sig}} / b_{\text{eff}}\,,
    \label{equ:fracDev}
\end{equation}

\noindent where \beff\ is the effective background below the signal peak. The number of effective background counts in a given energy window [$E_i^-, E_i^+$] is calculated as:


\begin{equation}
	b_{\text{eff}} = \frac{N}{\left( \sum_k \frac{ F^2(\Egamma)_{\text{sig},k} }{ F(\GamBkg)_{\text{bkg},k} } \right) -1}
       \label{eqn:beff}
\end{equation}

\noindent where the summation runs over 63 energy bins in each fit window, $N$ is the total number of events in the fit, $F_{\text{sig},k}$ and $F_{\text{bkg},k}$ are the 
binned probability distribution functions for the signal and background models: $F_{\text{sig},k} = \int_{\text{bin}~k} dE'~C_{\text{sig}}/n_{\text{sig}}$ and 
$F_{\text{bkg},k} = \int_{\text{bin}~k} dE'~C_{\text{bkg}}/n_{\text{bkg}}$, respectively. 
This definition of \beff\ is different than that presented in Sec. VIA of Ref.~\cite{REF:P7Line}. However both approximate the number of background events under the peak\footnote{\Eqref{eqn:beff} approximates the statistical uncertainty on \nSig\ such that $\delta\nSig\approx 1/\sqrt{\beff}$~(see Sec 3.1 of Ref.~\cite{Cowan:2010js}).}.  
By expressing \beff\ in terms of a sum of a function containing the signal and background models, it is straightforward to calculate this quantity for analyses beyond our line search. 
One could also expand the sum to include spatial bins as was done in a recent search for a DM signal in the Large Magellanic Cloud~\cite{Buckley:2015doa}. 
This quantity is especially useful when describing systematic uncertainties since \nSig\ and \beff\ for line-like features induced by systematic uncertainties are both expected to scale as $N$. The number of events in 
the energy window varies greatly across our energy ranges and ROIs.
In R180, the fit for \Egamma = 214 MeV has 38.9 million events, while the fit at \Egamma = 467 GeV in R3 has 52 events. The quantity $f$ in effect normalizes the large differences in the numbers of events. 
Note that the significance of a systematically induced line-like feature at a fixed $f$ will scale as $\sqrt{N}$.


Depending on the energy window, the fit will be dominated by either statistical ($\fstat = 1 / \sqrt{\beff}$) or systematic (\fsyst) uncertainties. 
It is critical to account for the systematic uncertainties in the fitting procedure since small line-like features ($f\sim0.01$) can be 
statistically significant, if \fsyst\ is not accounted for, when $\fstat \ll \fsyst$. For example, our fit at 947 MeV in R41 would have a 
local significance (see definition below) of $8.9\sigma$ if \fsyst\ were neglected. However, in that case $f=0.008$, which
is well within the systematic uncertainty range (see \secref{sec:syst}) and therefore cannot be deemed a detection.

If we increase the energy window width, the statistical uncertainty decreases because more events are introduced into the fit. 
However, the systematic uncertainty ($\delta n_{\text{syst}}=\fsyst\times\beff$) would then increase because the power-law approximation of the background energy spectrum becomes less valid. 
In our previous works we characterized the window width in terms of \sigE\ (68\% containment) at \Egamma. 
The 3.7-year search~\cite{REF:P7Line}, for which most fits were dominated by statistical uncertainties, used $\pm6\sigE$ windows. 
The 5.2-year low-energy search~\cite{REF:LELine} was dominated by systematic uncertainties and we found $\pm2\sigE$ windows to be optimal.
In order to have a common definition across our entire energy range, we chose to define the window width to vary with \Egamma: $\pm0.5\Egamma$. 
This causes the fit range to decrease naturally relative to 
$\sigE(\Egamma)$ at low energies and widen relative to $\sigE(\Egamma)$ at higher energies. 
A common window definition is useful since the size of line-like features observed in our control regions (which determines \fsyst) varies depending on the window width.

\subsection{Signal Significance and Trials Factor}\label{subsec:signif}

We define the local significance of a fit as the square root of the test statistic ($\slocal=\sqrt{\text{TS}}$), where the TS 
is defined as twice the logarithm of the ratio between the likelihood maximized for the signal hypothesis and for the null hypothesis (\nSig\ = 0):

\begin{equation}\label{eq:TS}
\text{TS}=2\textrm{ln}\frac{\mathcal{L}(\nSig = n_{\rm sig,best})}{\mathcal{L}(\nSig=0)}
\end{equation}

Since we perform many fits (121 fit energies in five ROIs), the local significance (\slocal) must be corrected by a trials factor to obtained the global significance (\sglobal). Since our fits largely overlap spatially and in energy, simply assuming all 605 fits are independent would severely overestimate our trials factor. Therefore, we determine the trials factor using background-only simulations. We follow the procedure outlined in Section VB of Ref.~\cite{REF:P7Line}. 
We created 1000 pseudo-experiments with our full-fit energy range for our five ROIs. The pseudo-experiments had a power-law energy spectrum with a \GamBkg = 2.3. We created five independent data sets at each energy which correspond to the spatially independent pieces of our nested ROIs. 
The number of events in each piece were taken from a Poisson randomization of the number of events seen in the actual data. We then combined the 
appropriate subsets to create the complete simulated data set for the whole region under study. 
For each pseudo-experiment (i.e. a simulation of one full search across the entire energy range and in all ROIs), we find the largest local significance (\smax). 
We can then use the cumulative distribution of~\smax\ to determine \sglobal\ for a given \slocal.
We show both the \smax\ distribution and the \slocal\ to \sglobal\ conversion for our line search in~\figref{TF}. We empirically found that the effective number of independent trials is $353\pm11$. 
In each of the five ROIs, the best fit number of independent fits is $\sim72.6$ taking into account the overlap in neighboring energy windows. 

\twopanel{ht}{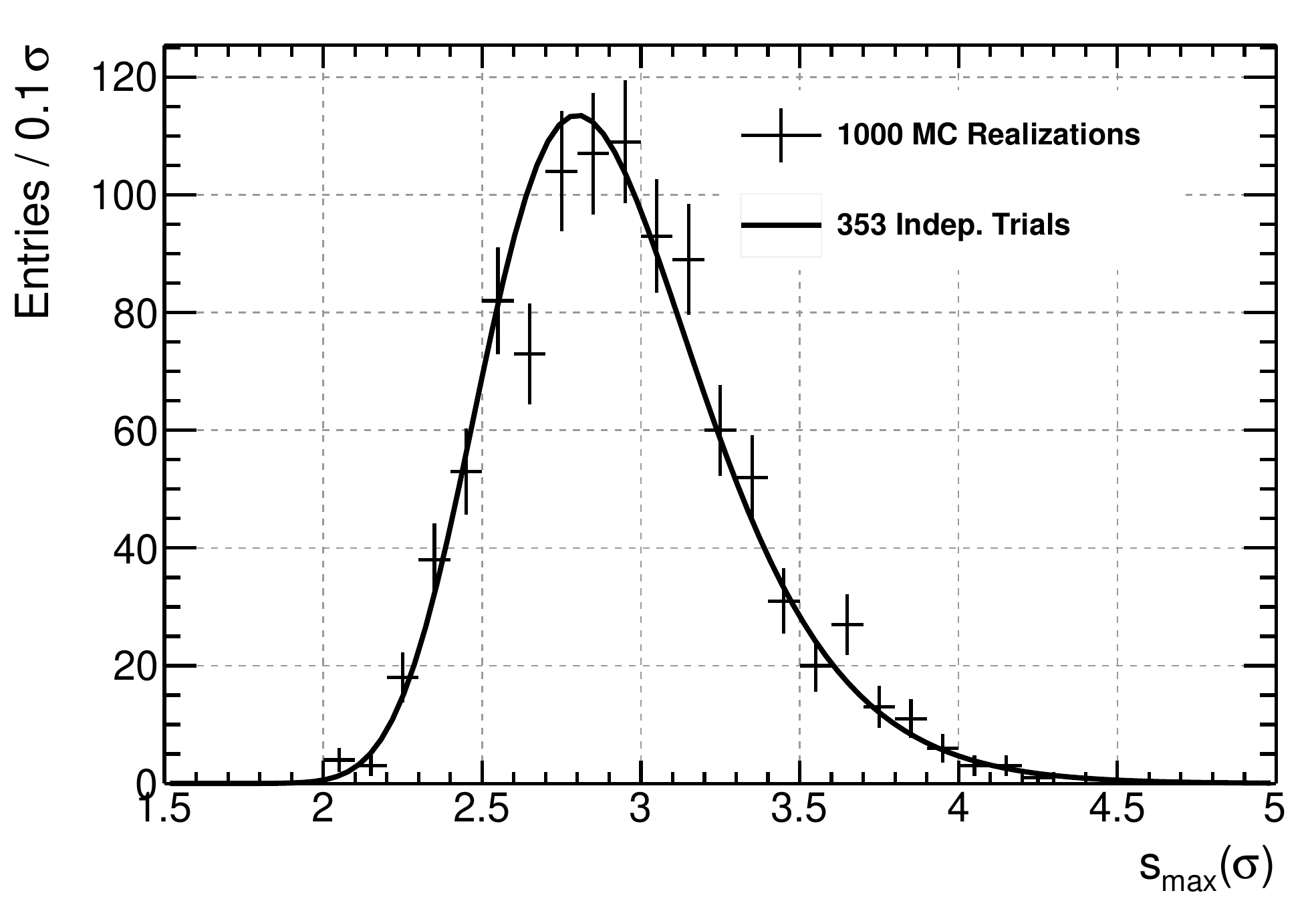}{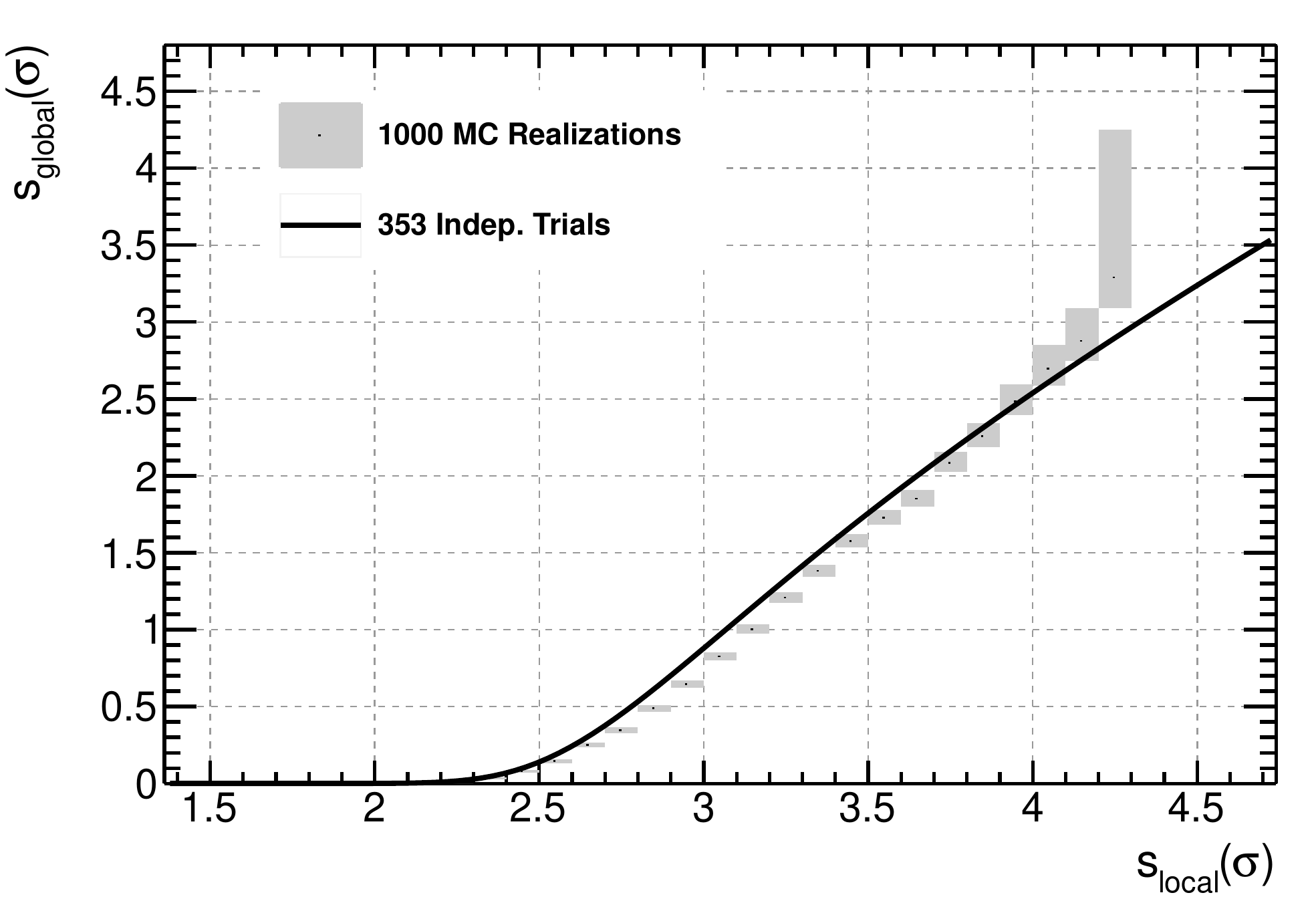}
{\caption{\label{fig:TF}Derivation of the global significance for a given local significance in this line search. Based on 1000 background-only pseudo-experiments of our full fit range. (Left) Distribution of the largest \slocal\ value (\smax) obtained in each pseudo-experiment. Each pseudo-experiment consisted of 605 fits. (Right) Corresponding \slocal\ to \sglobal\ conversion. }}

\section{SYSTEMATIC UNCERTAINTIES}\label{sec:syst}

To estimate the systematics, we perform scans for lines in control regions where
the background events vastly outnumber the signal events. 
Our main control region is the GP excluding the GC (the white region in \figref{ROIs}). 
Any line-like features detected along the GP would be induced by systematic uncertainties such as modeling deficiencies.

The two largest sources of systematic uncertainties are our modeling the background flux spectrum as a power law and our 
approximation of the energy-dependent variations in the exposure.
Since we do not mask point sources, the Galactic diffuse emission and point sources are considered together and approximated as a 
power law in each narrow energy window. Also, as was discussed in Ref~\cite{REF:P7Line}, fine energy-dependent variations 
in the LAT effective area are difficult to model accurately. Any discrepancy in the overall background model at \Egamma~(see~\Eqref{eq:BkgPDF_1D}) can be compensated by an excess or absorption-like feature.
We assume that the level of systematic uncertainties observed in the GP will also be present in our signal ROIs.

Following Ref.~\cite{REF:LELine}, we quantify the level of the systematic uncertainties by their fractional size \fsyst\ (see~\secref{sec:fitting} for details on $f$). 
In this way we can directly apply the \fsyst\ observed in the GP to our signal ROIs.
We scan for spectral lines from 200 MeV to 500 GeV in 31 $10^{\circ}\times10^{\circ}$ boxes along the GP. 
Specifically, we scan in regions where $|b|<5^{\circ}$ and $l > 35^{\circ}$ or $l < 325^{\circ}$. The results from this scan are shown in~\figref{GPfrac}.

\onepanel{ht}{\colSw{Figure_6}}{
\caption{\label{fig:GPfrac} Fractional signals ($f$) in the GP. Dots show observed $f$ in 31 $10^{\circ}\times10^{\circ}$ boxes along the GP. The solid red line is the average of the statistical uncertainties of the individual boxes. The blue dashed line is the value we chose to characterize \fsyst\ from modeling biases; see text for details.}}

The general trend of the fractional signal with energy shown in \figref{GPfrac} are due to the
interplay between the background model and energy-dependent variations in the
exposure, while the spread at each energy is due to variations from region to region in the astrophysical emission. To further investigate the energy-dependent variations in the exposure, we
analyzed two other control data sets: the Vela Pulsar and the Earth Limb (see
\appref{app:control_samples}). Note that the general energy-dependent behavior
of the observed fractional signals in each control region differs (see
\figref{control_fit_examples}). This is due to differences in the
deficiencies of our background models for each control region compared to the observed emission. 

\Figref{GPfrac} shows that the fits start to become dominated by statistical uncertainties ($\fstat>\fsyst$) around \Egamma~=~6~GeV. 
Therefore, we set $\delta f_{\text{GP}} = 0.015$ as the systematic level from the GP since it is the 68\% containment of $f_{\text{GP}}$ for $\Egamma < 5$ GeV. 
We note that the transition to dominance by statistical uncertainties is dependent on the total number of events in the fit, not \Egamma. 
For $\fsyst = 0.015$ this occurs around $N \sim10,000$. Additionally, the value of \fsyst\ we find here is slightly higher than that used in our 
5.2-year search~\cite{REF:LELine} since we are using wider energy windows (see~\secref{subsec:fitting}).

An additional potential source of systematic uncertainty is cosmic-ray contamination. To estimate $\delta f_{\text{CR}}$, we use the method described in Refs~\cite{REF:P7Line,REF:LELine}. 
Though a subdominant effect, cosmic-ray contamination is most significant at high Galactic latitudes. 
To estimate this uncertainty, we perform fits for lines using a sample of events in R180 that contains a high cosmic-ray contamination --- a background enriched (or ``dirty") sample. 
These are events that pass the \irf{P8\_SOURCE} selection, but not the \irf{P8\_CLEAN} selection. 
We then find the 68\% containment of the $f$ values observed in our background enriched sample to define $\delta f_{\text{CR,dirty}}$. 
To obtain the appropriate $\delta f_{\text{CR}}$ value to use in our \irf{P8\_CLEAN} signal dataset, we scaled $\delta f_{\text{CR,dirty}}$ using the \gammaRayHyph\ acceptance ratio between \irf{P8\_CLEAN} and \irf{P8\_SOURCE} and the observed number of events in both event selections. See App. D5 in Ref~\cite{REF:P7Line} for more details. 
We estimate $\delta f_{\text{CR}}\sim$  0.01 in R180 and R90. In R41, R16, and R3, $\delta f_{\text{CR}}$ is negligible ($<0.003$). 
Adding $\delta f_{\text{CR}}$ in quadrature with $\delta f_{\text{GP}}$ gives the total systematic uncertainty in each ROI: \fsyst = 0.016 for R180 and R90 and \fsyst = 0.015 for R41, R16, and R3.


\section{FITTING RESULTS}\label{sec:results}
In our search for \gammaRayHyph\ spectral lines in five ROIs, the data did not yield any globally significant lines. 
Our most significant fit occurred at 115 GeV in R16 and had a local significance of $2.8\sigma$, which corresponds to a global significance of $0.4 \sigma$ (see \figref{TF}).
In the case of a null result, the local fit significance will follow a one-sided Gaussian according to Chernoff's theorem~\cite{REF:Chernoff}: half the local fit significances will be zero since we require $\nSig>0$. 
In \figref{SignifDist}, we fit our local significance distribution to a one-side Gaussian function and find a best-fit width of $0.81\pm0.03$. This is close to the expected value of one; however, 
a width less than one suggests our \fsyst\ determined a priori is a slight overestimation.

\onepanel{ht!}{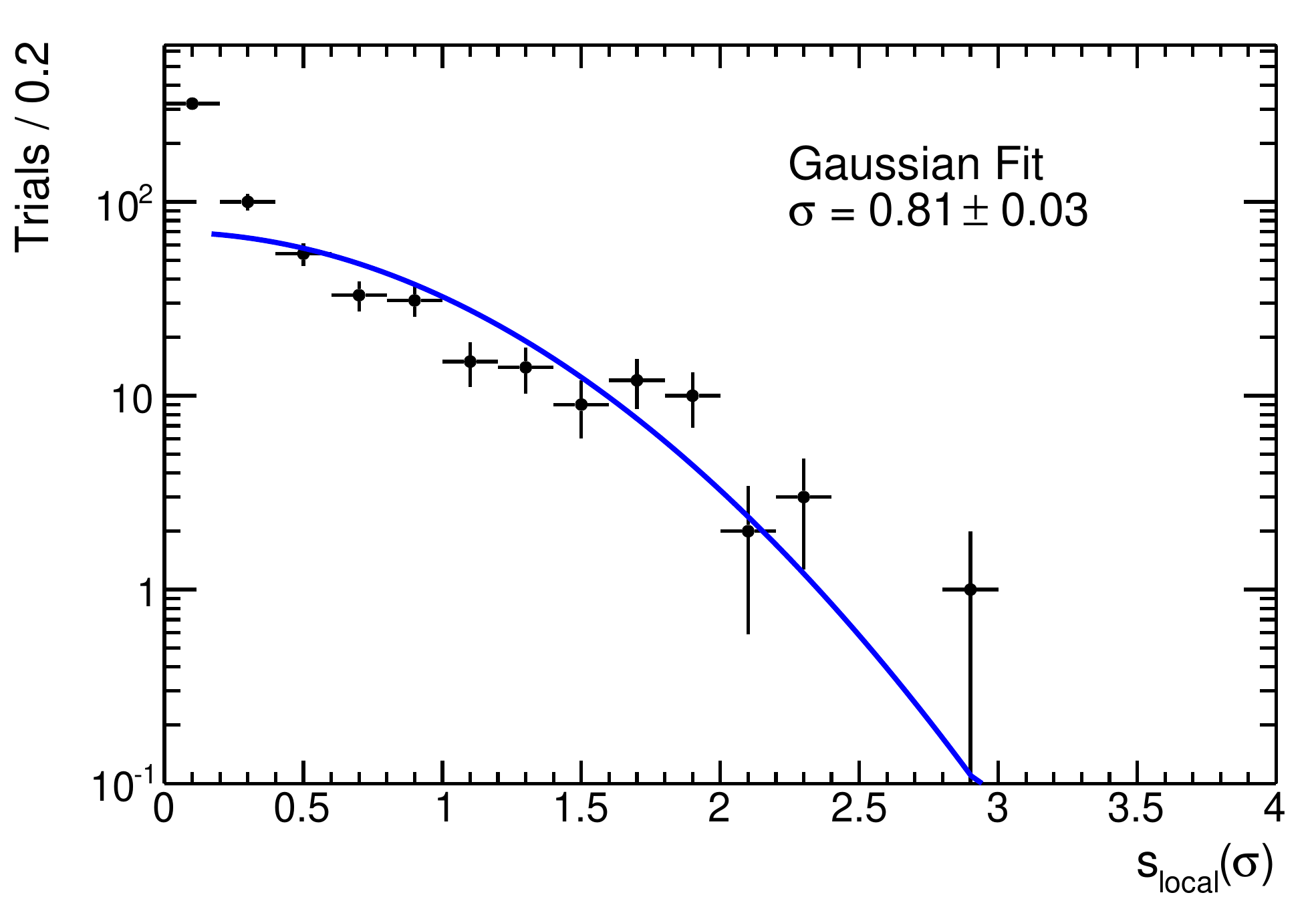}{\caption{\label{fig:SignifDist}Distribution of local significance values for all our fits. Best fit to a one-sided Gaussian is shown.}}

Since no significant spectral lines are detected, we set flux upper limits for monoenergetic emission. We also derive limits for DM annihilation and decay from our flux limits. We obtain a 95\% confidence level (CL) counts upper limit by increasing \nSig\ until the logarithm of the likelihood decreased by 1.36 (2.71/2) with respect to the maximum. Using the average exposure in each ROI at \Egamma\ ($\Exposure_{ROI}(\Egamma)$), we can find the 95\% CL monoenergetic flux upper limit using:

\begin{equation}\label{equ:FluxToCnt} 
\PhiMono(\Egamma)=\frac{\nSig(\Egamma)}{\Exposure_{ROI}(\Egamma)}\,.
\end{equation}

\noindent If the spectral line is produced by DM annihilation directly into a pair of \gammaRays\ the expected differential flux is given by:

\begin{equation}\label{eq:annSigTot}
  \left(\frac{d\Phi_\gamma}{dE}\right)_{\rm ann} = \frac{1}{8\pi} \, \frac{\AnnGG}{m_{DM}^2}\,
\left(\frac{dN_\gamma}{dE}\right)_{\rm ann} \intROI \frac{dJ_{\rm ann}}{d\Omega}~d\Omega
\end{equation}

\noindent where $m_{DM}$ is the mass of the DM particle, $\left(\frac{dN_\gamma}{dE}\right)_{\rm ann} = 2\delta(\Egamma-\Ereco)$, and $\Egamma = m_{DM}$. For lines produced by DM decay into a \gammaRay\ and a second neutral particle, the expected flux is given by:

\begin{equation}\label{eq:decSigTot}
\left(\frac{d\Phi_\gamma}{dE}\right)_{\rm decay} = \frac{1}{4\pi} \,\frac{1}{\tau_{DM}}\,\frac{1}{m_{DM}}\,
\left(\frac{dN_\gamma}{dE}\right)_{\rm decay} \intROI \frac{dJ_{\rm decay}}{d\Omega}~d\Omega
\end{equation}

\noindent where $\tau_{DM}$ is the DM lifetime, $\left(\frac{dN_\gamma}{dE}\right)_{\rm decay} = \delta(\Egamma-\Ereco)$, and $\Egamma = m_{DM}/2$. 
The ``J factors" ($J_{\rm ann/decay}$) are proportional to the expected intensity of \gammaRayHyph\ emission from DM annihilation or decay in a given ROI assuming a specific DM density distribution $\rho(r)$. They are defined as an integral over the line of sight of the DM density:

\begin{equation}\label{eq:J_ann}
\frac{dJ_{\rm ann}}{d\Omega} = \int_{l.o.s.} ds~ \rho(r)^2
\end{equation}

\noindent and 

\begin{equation}\label{eq:J_decay}
\frac{dJ_{\text{decay}}}{d\Omega} = \int_{l.o.s.} ds~ \rho(r).
\end{equation}

Using the $\Phi_{mono}$ 95\% CL upper limits derived using \Eqref{equ:FluxToCnt}, we solve \Eqref{eq:annSigTot} and \Eqref{eq:decSigTot} for the 95\% CL upper limits on \AnnGG\ and the 95\% CL lower limits on $\tau_{DM}$ respectively. The flux limits and DM limits are given in Tabs.~\ref{tab:allLimits0}--\ref{tab:allLimits2}.

\Figref{SigmaVUL} shows the \AnnGG\ 95\% CL upper limits in our four
ROIs optimized for sensitivity to DM annihilation and \figref{TauUL} shows the $\tau_{DM}$ lower limits in R180. Also shown are the
corresponding limits from our previous 3.7-year analysis~\cite{REF:P7Line} and
our previous 5.2 year analysis~\cite{REF:LELine}. Two main factors contribute to
the differences in these three sets of limits: different depths of exposure, and
different approaches for the treatment of systematic uncertainties. 
As was discussed in \secref{sec:method_data}, while the acceptance of the LAT increased in \irf{Pass 8}, the energy resolution did not significantly improve. The results for each ROI benefited from the increased exposure due to the larger effective area in \irf{Pass 8}. Also, our smallest ROIs (R3 and R16), benefited from the increased exposure of the GC region during the 6th year of data taking: from December 4th, 2013 to December 4th 2014, \Fermi\ operated in a modified observing mode\footnote{\url{http://fermi.gsfc.nasa.gov/ssc/proposals/alt_obs/obs_modes.html}} that roughly doubled the rate of increase of exposure in the GC relative to normal survey mode. 

The 3.7-year analysis did not incorporate systematic uncertainties into calculating the limits. As was shown in Ref~\cite{REF:LELine}, accounting for systematic uncertainties makes the results more robust, 
especially for fits with a large number of events where the systematic uncertainties dominate. In our 5.2-year analysis, we chose a conservative \fsyst\ value that resulted in all of the fits having a local significances less than $1\sigma$. In this work, we used a more realistic \fsyst\ value, which results in a distribution of the local fit significances that is significantly closer to a one-side Gaussian function (see \figref{SignifDist}). 
Therefore, on average, our current limits should represent a greater improvement over the 5.2-year results than would be expected solely from the increased exposure, since the 5.2-year analysis was more conservative.

\fourpanel{ht}{\colSw{Figure_8a}}{\colSw{Figure_8b}}{\colSw{Figure_8c}}{\colSw{Figure_8d}}{\caption{\label{fig:SigmaVUL}95\%
CL $\langle\sigma v\rangle_{\gamma\gamma}$ upper limits for each DM profile considered in the corresponding optimized ROI. The upper left panel is for the NFWc ($\gamma$=1.3) DM profile in the R3 ROI. 
The discontinuity in the expected and observed limit in this ROI around 1 GeV is the result of using only PSF3 type events. See~\secref{sec:method_roi}~for more information. 
The upper right panel is for the Einasto profile in the R16 ROI. The lower left panel is the NFW DM profile in the R41 ROI, and finally the lower right panel is the Isothermal DM profile in the R90 ROI.
Yellow (green) bands show the 68\% (95\%) expected containments derived from 1000 no-DM MC simulations (see~\secref{subsec:signif}).  The black
dashed lines show the median expected limits from those simulations. Also shown are the limits obtained in our 3.7-year line search~\cite{REF:P7Line} and our 5.2-year line search~\cite{REF:LELine} when the assumed DM profiles were the same.}} 

\onepanel{ht}{\colSw{Figure_9}}{\caption{\label{fig:TauUL}95\%
CL $\tau_{DM}$ lower limits assuming and NFW profile in R180. Yellow (green) bands show the 68\% (95\%) expected containments derived from 1000 no-DM MC simulations (see~\secref{subsec:signif}).  The black dashed lines show the median expected limits from those simulations. Also shown are the limits obtained in our 3.7-year line search~\cite{REF:P7Line} and our 5.2-year line search~\cite{REF:LELine}}}

\section{THE LINE-LIKE FEATURE NEAR 133~GeV}\label{sec:133GeV_Feature}

The \gammaRayHyph\ spectrum in the energy range near 133~GeV has been of particular interest after a potential signal was reported based on 3.7 years of data for a small ROI containing the GC region~\cite{REF:Weniger:2012tx,Bringmann:2012vr}.
A similar, yet not globally significant, feature was also reported by the LAT Collaboration~\cite{REF:P7Line}. 
There have been two relevant developments since the previous results. The first is greater exposure toward the GC, due in part to the modified observing strategy described in~\secref{sec:results}. 
The second is the implementation of the \irf{Pass 8} event classification that we use here (see~\secref{sec:method_data}). 
In this section we first compare the overlapping events between \irf{Pass 7REP}~and \irf{Pass 8}~(\secref{subsec:LineEvtLevel}), and then measure how the apparent signal has evolved with additional data in both R3 (\secref{subsec:LineR3}) and the Earth Limb (\secref{subsec:LineLimb}). 

\subsection{Event-Level Comparison of \irf{Pass 7REP} and \irf{Pass 8}}\label{subsec:LineEvtLevel}

We first compare the reconstructed energies of events that are in both the \irf{P7REP\_CLEAN} and \irf{P8\_CLEAN} event classes (or event selections) and in the smallest ROI (R3). 
Events must pass the event selections as outlined in~\tabref{selection}, be located in R3, and must have measured energies in \irf{Pass 7REP} greater than 20 GeV to be considered for this comparison. 
A comparison of these events and events with energies between 120--150 GeV~is shown in the left panel of~\figref{EDispP7P8}. 
The distribution of energy differences between \irf{Pass 7REP}~and \irf{Pass 8}~is similar in shape for events in the window around 133 GeV as for all events above 20 GeV. 
We made similar studies in the ROIs outlined in~\secref{sec:method_roi}, in the Earth Limb, and found similar results. 

The LAT consists of 16 towers, each includes a tracker module and a calorimeter module~\cite{REF:2012.P7Perf}. \irf{Pass 8} includes important updates to the 
energy reconstruction near the edges of the calorimeter modules ($<$60 mm from the center of the gap)~\cite{Atwood:2013rka,Bruel:2012bt}. 
Events that deposit the majority of their energy (or have their reconstructed centroid) near the edge of a calorimeter module are more difficult to reconstruct accurately because of energy leakage of the shower into the 
gaps between modules, or towers. \irf{Pass 8}~applies an improved handling of this leakage in the energy reconstruction algorithms. 
We show in~\figref{EDispP7P8} the distance of each reconstructed centroid from the center of the calorimeter gap for the events passing the comparison selection outlined above. 
Each calorimeter crystal has
a width of 326 mm and the gap between modules of 44 mm~\cite{REF:2009.LATPaper}. This yields a total width of 370 mm. In this figure, 0 mm marks the 
distance from the middle of the gap between sets of crystals. The figure at the top also includes a cartoon to illustrate the location of the edge of the calorimeter crystal with the 
center located at 185 mm. 

\twopanel{ht}{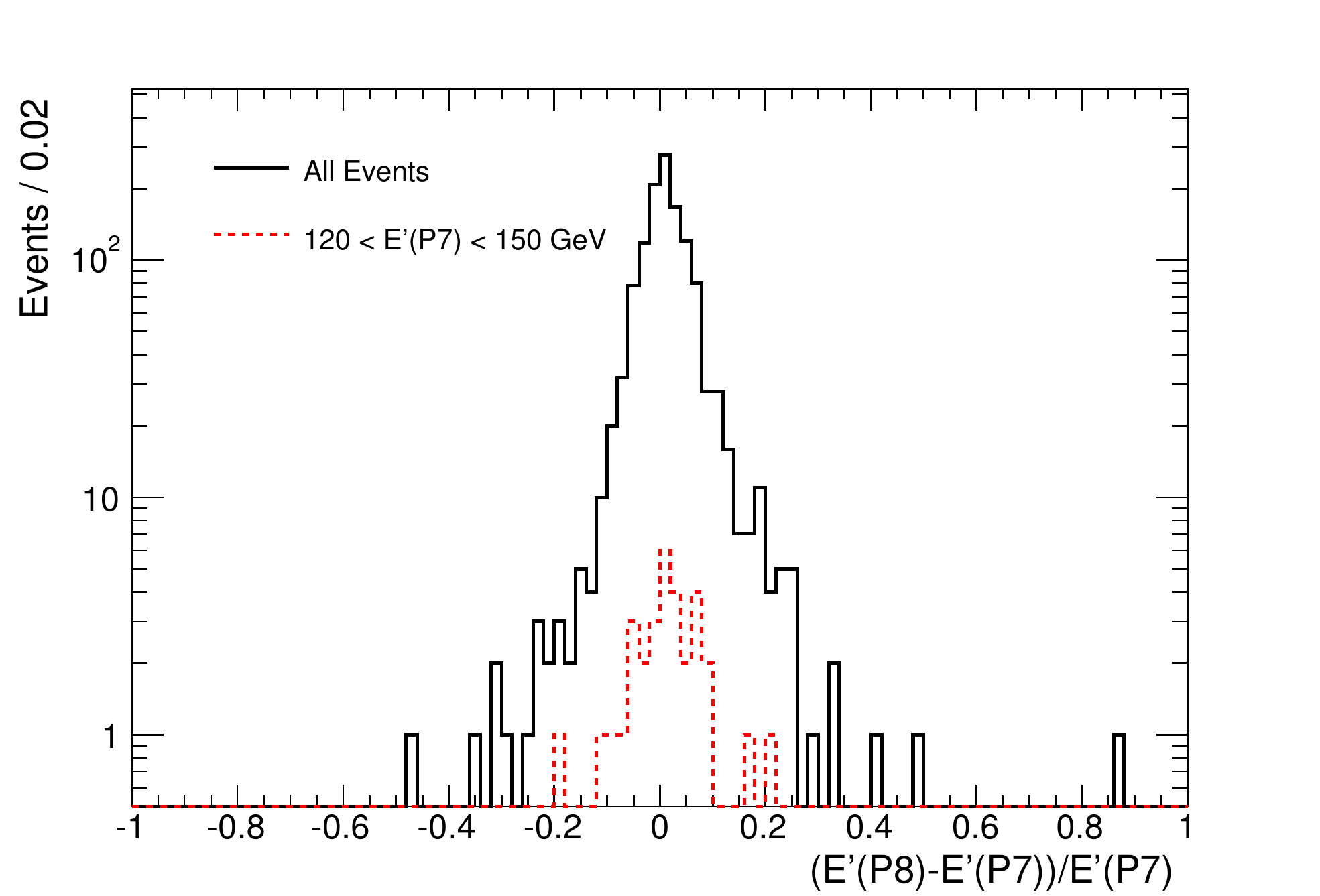}{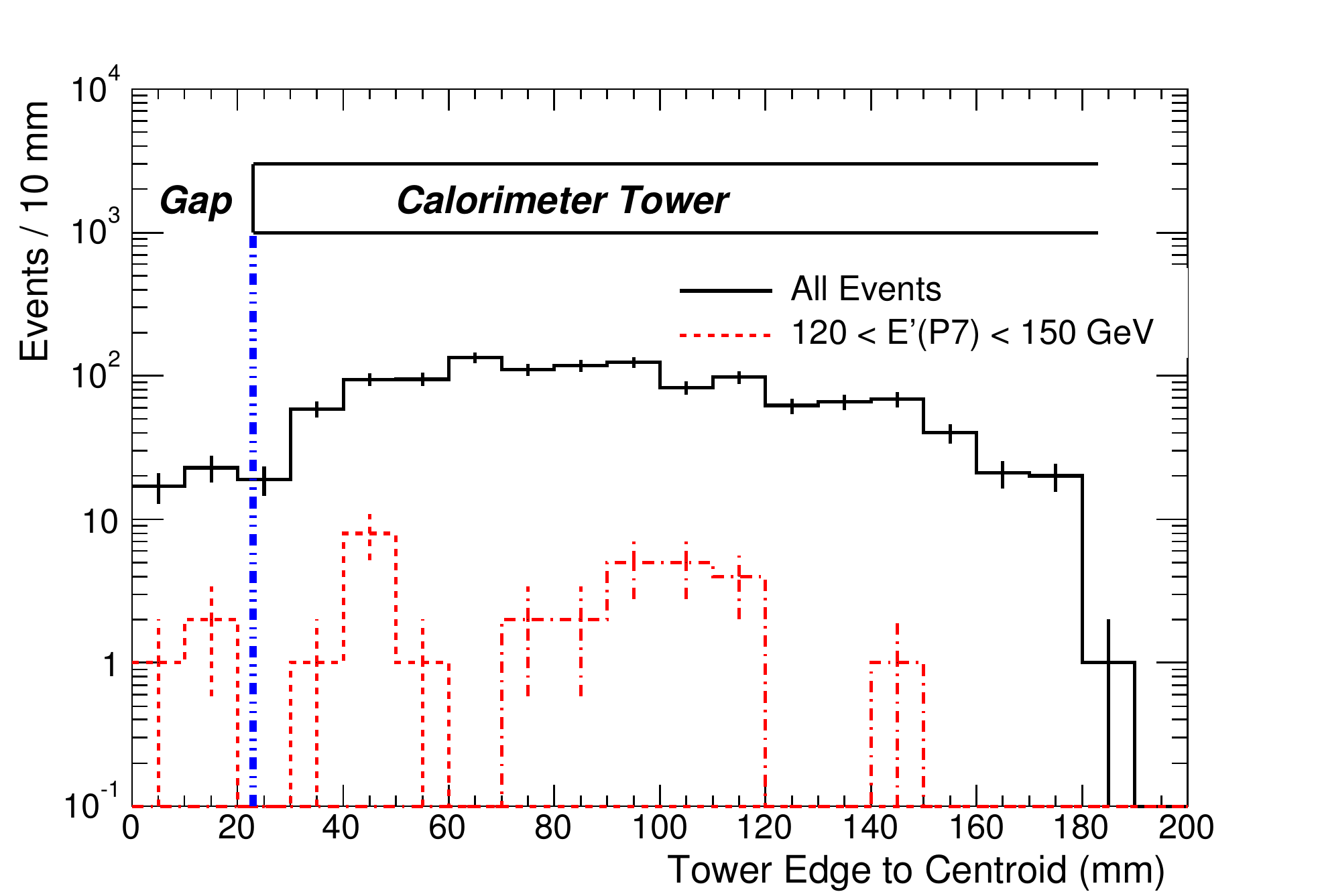}
{\caption{\label{fig:EDispP7P8} Comparison of events selected in R3 from the full data set (5.8 years) passing the Clean event selections simultaneously in both \irf{Pass 7REP} and \irf{Pass 8}. 
The left panel shows the fractional energy difference between \irf{Pass 7REP} and \irf{Pass 8}. 
The black line shows all events that have a reconstructed energy (\Ereco) above 20 GeV. The dashed red line indicates events which are 
with \irf{Pass 7REP} energies in the range 120--150 GeV. 
The right panel shows the distribution of the distance from the center of the gap of the calorimeter module to the centroid of the cluster. The black curve shows all events above 20 GeV 
and the red curve has the subset in the \irf{Pass 7REP} range 120--150 GeV. These events are separated by those near the edge of the tower ($<$60 mm from the center of the gap), 
and those away from the edge of the tower ($>$60 mm from the center of the gap). The blue dashed line marks the edge of the calorimeter module, which is also illustrated at the top of the figure.}}

About half of the overlapping events between \irf{Pass 7REP}~and \irf{Pass 8}~in the 120--150 GeV energy range were reconstructed with centroids near the edges of the towers ($<$60 mm from the center of the gap). 
As a consequence, these events had the largest differences in reconstructed energy and comprised the tails of the distribution shown on the left in~\figref{EDispP7P8}. 
There appears to be a slight enhancement of events where much of the shower was lost between modules in the energy range around 133~GeV relative to all events above 20~GeV.

\subsection{Feature in R3}\label{subsec:LineR3}

To understand the impact of \irf{Pass 8} on R3, 
we first considered the same time and energy range as our previous 3.7-year search~\cite{REF:P7Line}. The feature in \irf{Pass 7REP}, which was narrower than the 
energy resolution of the LAT and had a local significance of 3.3$\sigma$. With \irf{Pass 8} 
the excess present in \irf{Pass 7REP}~data is reduced to a local significance of 2$\sigma$ as is shown in 
\figref{37Line}. 

\onepanel{ht}{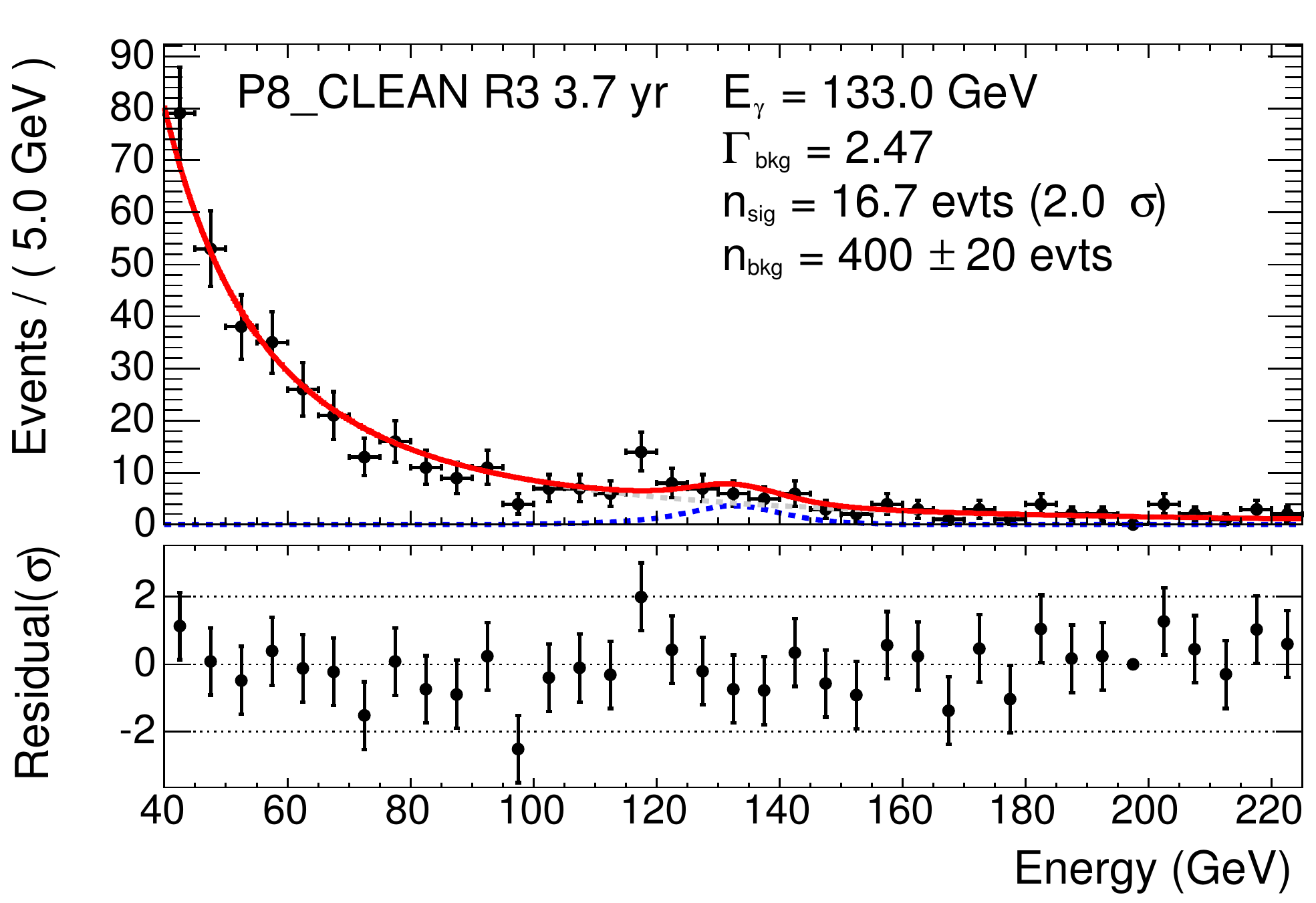}
{\caption{\label{fig:37Line} Fit at 133 GeV for a \gammaRayHyph~in the 3.7-year \irf{Pass 8} data set using the 2D energy dispersion model in R3. 
The solid curve shows signal and background fitting procedure described in \secref{subsec:fitting}. The blue dotted line is the signal model that best fits the data. The gray line, which is mostly 
hidden by the solid curve, is the best fit background. The bin 
size is such that the energy resolution is sampled with 3 bins. 
}}

We then considered the data for the full 5.8-year time range. \Figref{Line} shows 
the fit for a \gammaRayHyph~line at 133 GeV in the 5.8 year \irf{Pass 7REP}~and \irf{Pass 8}. 
The \irf{Pass 8} data are fit using the method described in \secref{sec:fitting} and the \irf{Pass 7REP}
data are fit similarly, but using the `2D' $D_{\text{eff}}$ model described in Sec.~IV of Ref.~\cite{REF:P7Line}. 
The \irf{Pass 7REP}~curve shows a clear decrease 
in local significance (from 3.3$\sigma$ to 2$\sigma$) with respect to the previous line analysis over a shorter time interval~\cite{REF:P7Line}. Similarly for the 
\irf{Pass 8}~data, the local significance also decreases (2$\sigma$ to $<$1$\sigma$) using the full 5.8 year dataset.

\twopanel{ht}{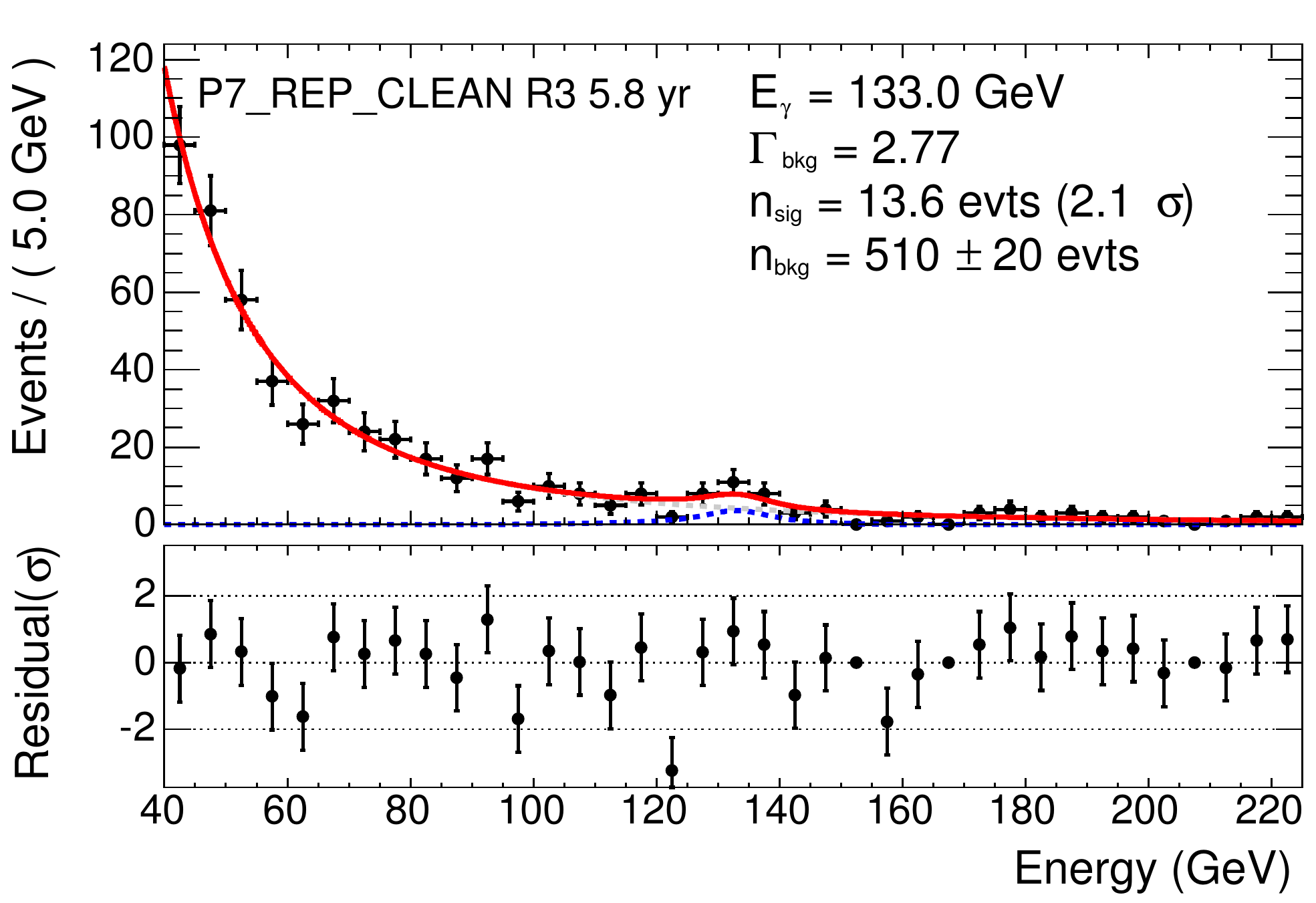}{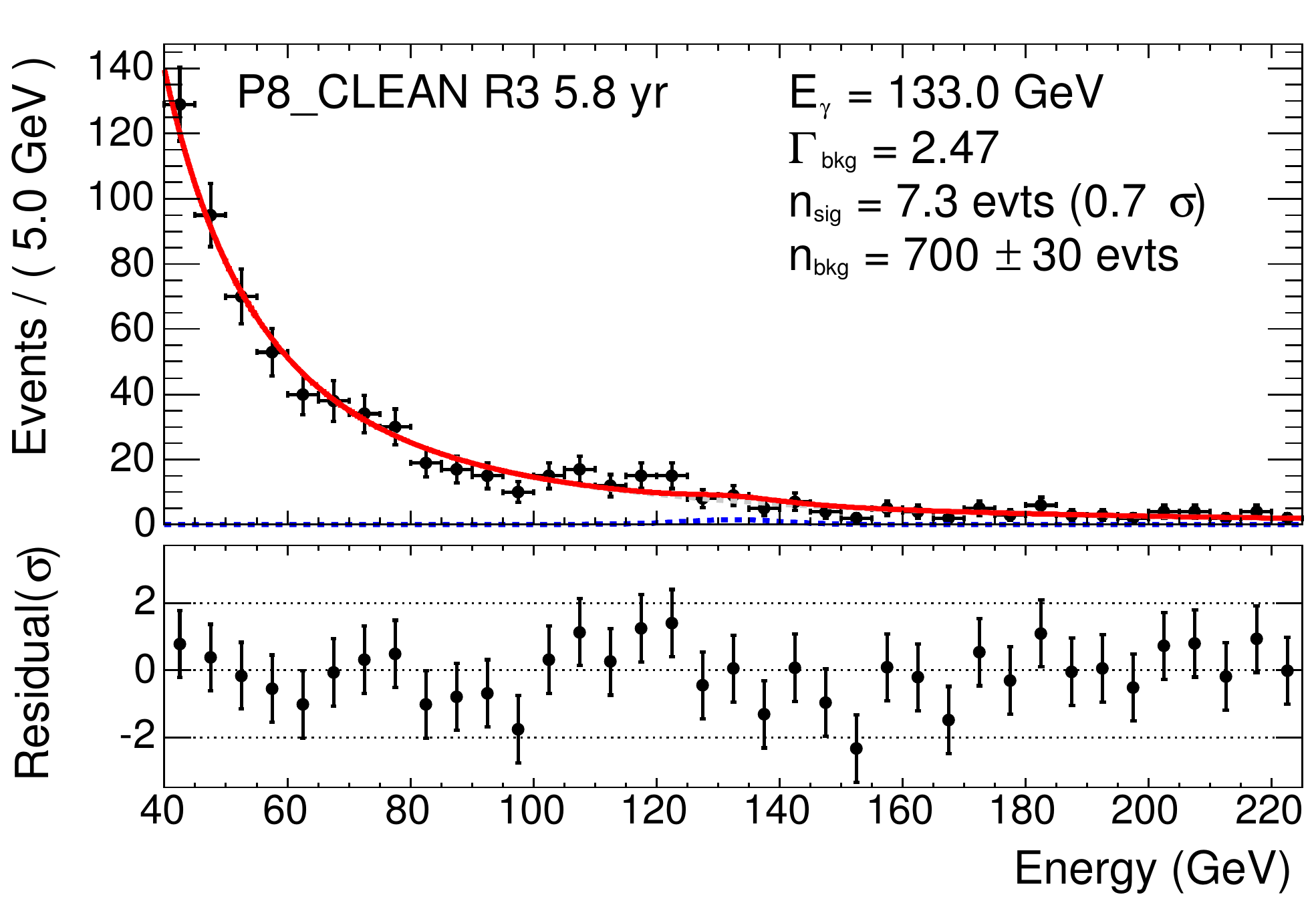}
{\caption{\label{fig:Line} 
Fit at 133 GeV for a \gammaRayHyph~in a 5.8-year \irf{Pass 7REP}~(left) and \irf{Pass 8}~(right) data sets using the 2D energy dispersion model in R3. 
The solid curve shows signal and background fitting procedure described in \secref{subsec:fitting}. The blue dotted line is the signal that best fits the data. The gray line, which is mostly 
hidden by the solid curve, is the best fit background. The bin 
size is such that the energy resolution is sampled with 3 bins.}}

\subsection{Feature in the Earth Limb}\label{subsec:LineLimb}

The \gammaRayHyph~spectrum of the Earth Limb (see \tabref{selection}) is expected to be featureless; however, 
in the \irf{Pass 7REP}~data a 2$\sigma$ feature was found at the same energy as the feature in R3~\cite{REF:P7Line, Ackermann:2014ula}. This was a strong indication that the 
feature seen in R3 could have been, in part, a systematic effect. We carried out additional studies with \irf{Pass 8}~event reconstruction and the full dataset to further 
understand this feature in the Limb. 
\Figref{limb}~shows a fit to a \gammaRayHyph~line at 133 GeV using the full 5.8 year \irf{Pass 7REP}~and \irf{Pass 8}. 
We find a slight detection of a line-like feature in both \irf{Pass 7REP}~and \irf{Pass 8} with a similar fractional size. 
With \irf{Pass 8}~the significance increases slightly due mainly to the increase in the number of events from the greater acceptance of \irf{Pass 8}. 

\twopanel{ht}{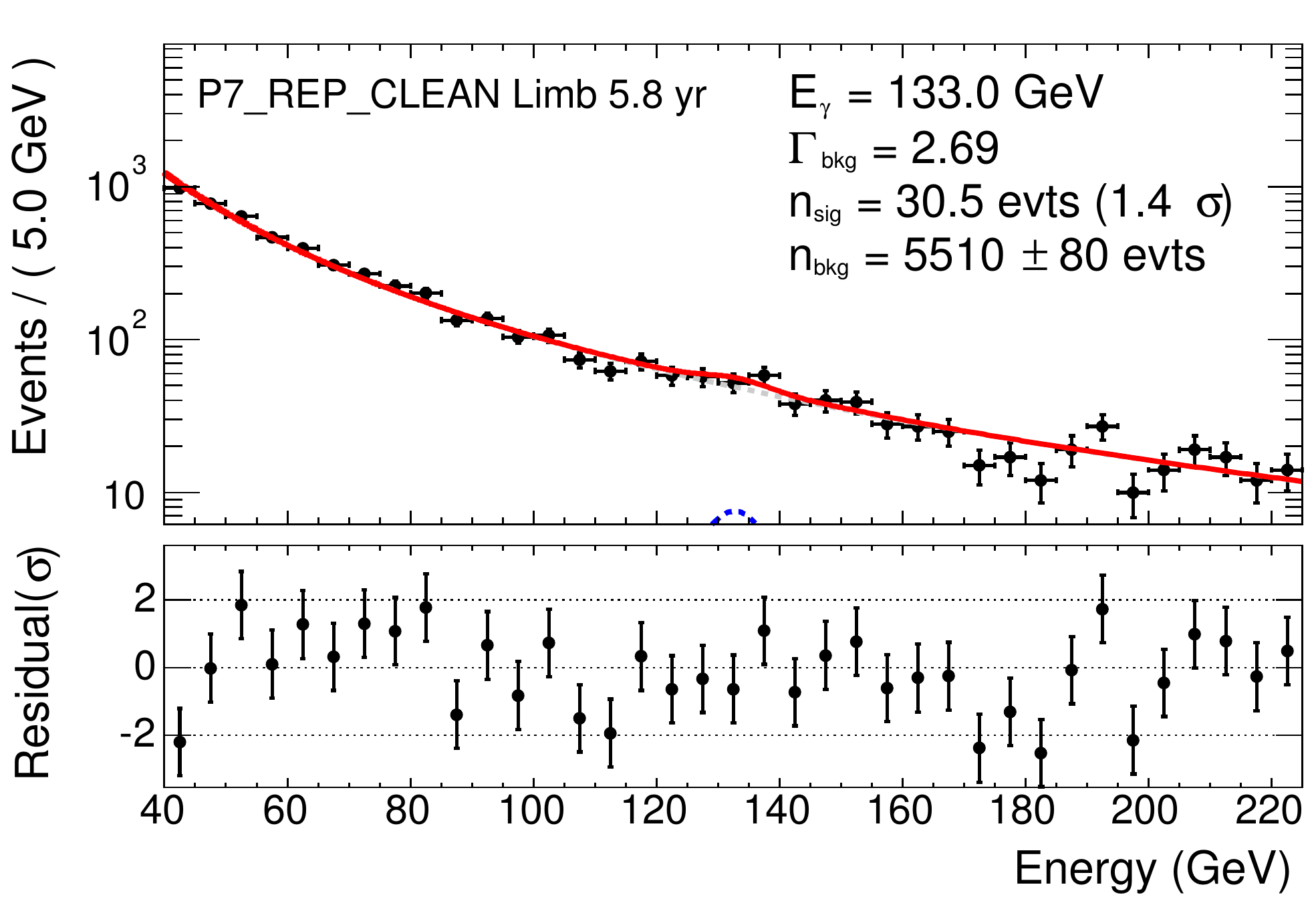}{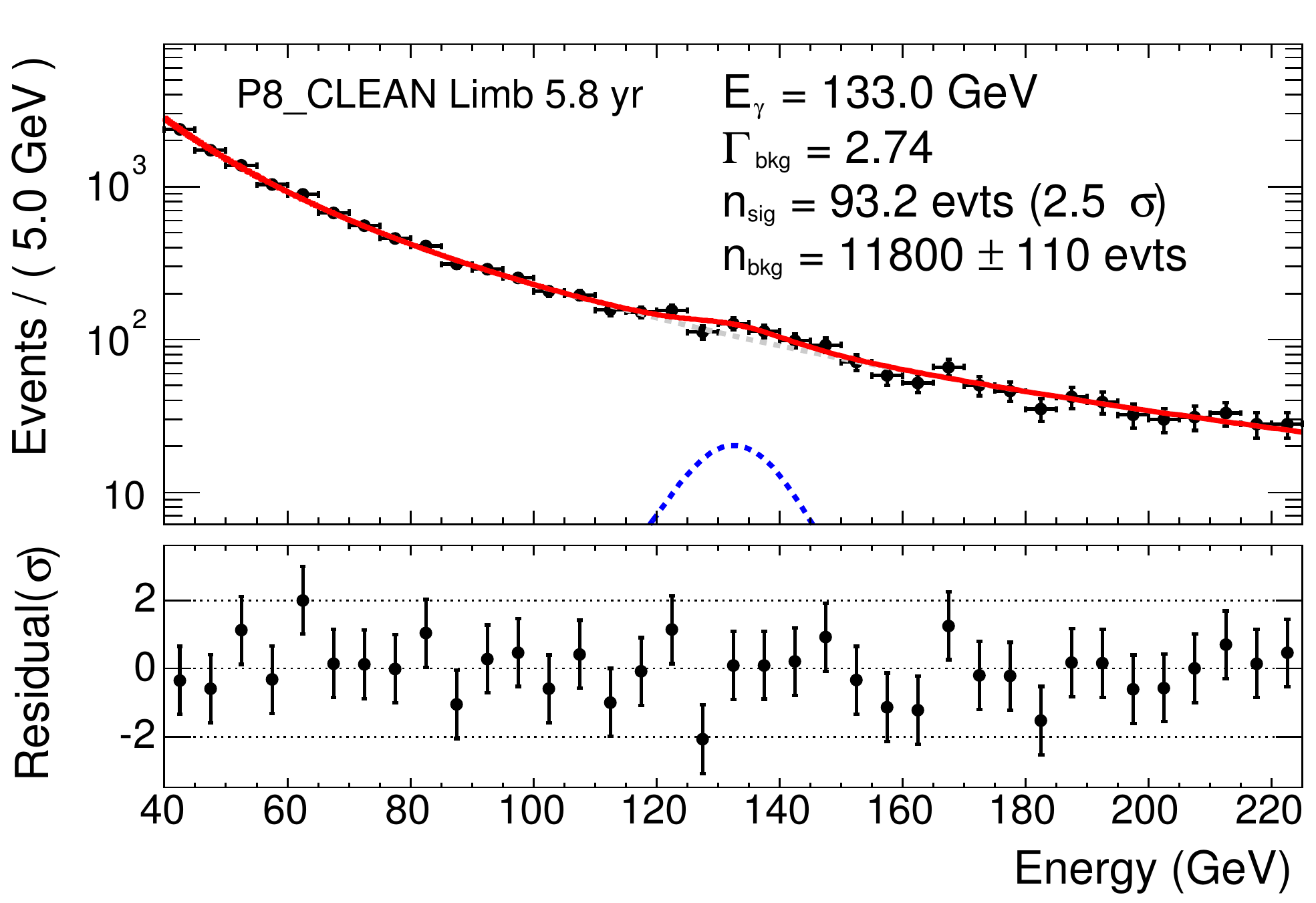}
{\caption{\label{fig:limb} 
Fit at 133 GeV for a \gammaRayHyph~in a 5.8-year \irf{Pass 7REP}~(left) and \irf{Pass 8}~(right) data sets using the 2D energy dispersion model in the Earth's Limb. 
The solid curve shows signal and background fitting procedure described in \secref{subsec:fitting}. The blue dotted line is the signal that best fits the data. The gray line, which is mostly 
hidden by the solid curve, is the best fit background. The bin 
size is such that the energy resolution is sampled with 3 bins.}}

We note that no feature at 133 GeV is present in the GP control region (\secref{sec:method_roi}). 
To try to understand the nature of the slight excess in the Earth Limb with no detection in the GP, events in the GP were reweighted 
in $\theta$ and in azimuthal angle, $\phi$, to the distribution in the Limb. 
This would indicate a dependence of  the feature on the particular distribution of arrival directions of the \gammaRays\ in instrument coordinates. 
The reweighting also yielded no detection of a line-like feature at 133 GeV in the GP. Additionally the Limb selection criteria was 
modified (in both $\theta_{r}$ and $\theta_{z}$) to see if the feature was enhanced or decreased in any particular part of phase space. The only significant change came when splitting the 
Limb data by the signed value of the rocking angle, $\theta_{r}$. The feature appears more significantly (2.6$\sigma$) for time intervals when the rocking angle of the LAT was positive 
($\theta_{r}>$52$^{\circ}$), and almost disappears (0.75$\sigma$) during time intervals when the rocking angle was negative ($\theta_{r}<-$52$^{\circ}$). 
Requiring larger values of $|\theta_r|$, however, does not significantly change the fractional signal or the significance.

\section{SUMMARY}\label{sec:summary}

In this work, we have presented an updated search for \gammaRayHyph\ spectral lines using techniques developed in our previous line searches~\cite{REF:2010.LineSearch,REF:2012.LineSearch,REF:P7Line,REF:LELine} 
across more than three decades of energy using data reprocessed and selected with \irf{Pass 8}. 
We searched for spectral lines in the energy range from 200 MeV to 500 GeV in five ROIs optimized for 
signals originating from Galactic DM annihilation or decay.  We do not find any significant spectral lines and therefore 
set monoenergetic flux upper limits in each of our ROIs.   

Our search improves on our most recent 3.7-year analysis~\cite{REF:P7Line} and 5.2-year analysis~\cite{REF:LELine} with increased exposure and a broader energy range. 
This results in a corresponding increase in the exposure for the entire 5.8-year time range.  Additionally, the exposure toward our smallest ROIs (R3 and R16) benefited 
from \Fermi\ operating in a modified observing mode that roughly doubled the rate of accumulation of exposure in the GC region for the last five months of the dataset 
relative to normal survey mode.

Additionally, we have improved our treatment of the systematic uncertainties in this analysis. We explicitly incorporate the systematic uncertainties 
in our limits by including a nuisance parameter in our likelihood function using an \fsyst\ value based on fits for line-like features along the GP. 
Our current limits improve, on average, relative to our 5.2-year limits in part because we use a more realistic \fsyst\ value. In fits for which systematic 
uncertainties were the dominant uncertainty (i.e. lower energy fits in larger ROIs that had more than $\sim$10,000 events\footnote{For example, in our search, fits with \Egamma $<$ 10 GeV and \Egamma $<$ 150 GeV have \fsyst $>$ \fstat\ in R3 and R180 respectively.}), our 5.8-year limits are more robust.

We investigated in particular a previously-reported line-like feature at 133 GeV after an initial excess was found with local significance 3.3$\sigma$ ~\cite{REF:P7Line}. Two additional years
of data and new event reconstruction and selection algorithms see the feature decrease from $f(133~\rm{GeV})_{\rm{R3}}=0.61$ in the 3.7-year \irf{Pass 7REP} data set to $f(133~\rm{GeV})_{\rm{R3}}=0.07$ in the 5.8-year~\peight\ dataset. This fractional signal in~\peight\ is consistent with what is also seen in the Earth Limb control region. With the entire 5.8-year~\peight\ data set, the local significance has dropped to 0.72$\sigma$ (from 2.0$\sigma$ in the 3.7-year data set), which is consistent with most of the original feature originating from a statistical fluctuation. The fact that the feature at 133 GeV is still marginally significant in the Earth Limb suggests a small systematic effect at this energy; however, no such feature is present in the GP control data set. 

The sensitivity of future line searches with the LAT will increase with continued exposure. Additional improvements require a more sophisticated modeling of the standard astrophysical backgrounds beyond our simple power-law approximation.

\begin{acknowledgments}
The \textit{Fermi}-LAT Collaboration acknowledges generous ongoing support
from a number of agencies and institutes that have supported both the
development and the operation of the LAT as well as scientific data analysis.
These include the National Aeronautics and Space Administration and the
Department of Energy in the United States, the Commissariat \`a l'Energie Atomique
and the Centre National de la Recherche Scientifique / Institut National de Physique
Nucl\'eaire et de Physique des Particules in France, the Agenzia Spaziale Italiana
and the Istituto Nazionale di Fisica Nucleare in Italy, the Ministry of Education,
Culture, Sports, Science and Technology (MEXT), High Energy Accelerator Research
Organization (KEK) and Japan Aerospace Exploration Agency (JAXA) in Japan, and
the K.~A.~Wallenberg Foundation, the Swedish Research Council and the
Swedish National Space Board in Sweden.

Additional support for science analysis during the operations phase is gratefully
acknowledged from the Istituto Nazionale di Astrofisica in Italy and the Centre National d'\'Etudes Spatiales in France.

\end{acknowledgments}

\clearpage

\appendix
\section{OTHER CONTROL SAMPLES}\label{app:control_samples}

Our main quantification of \fsyst\ was determined based on fits in $10^{\circ}\times10^{\circ}$ ROIs along the GP. This is because the background \gammaRayHyph\ emission in the GP is broadly similar to that in our signal ROIs. Systematically-induced line-like features in the GP could be produced by approximating the background spectrum as a power law and unmodeled energy-dependent variations in the exposure. 

One can also use other control samples to further study systematically-induced line-like features. The two additional control samples we
consider are the Earth's Limb and the Vela Pulsar.  Both are extensively used by the LAT Collaboration for studies of systematic uncertainties~\cite{REF:2012.P7Perf} and were also used as control samples in previous line searches \cite{REF:P7Line,REF:LELine}. It should be noted that the intrinsic \gammaRayHyph\ emission in these two control regions is different from that in the GP.  

The Earth's Limb emission is composed of \gammaRays\ produced via CR interactions in the Earth's upper atmosphere.  The Limb dataset is composed of events which pass the selection $|\theta_r|>52^{\circ}~\&~111^{\circ} < \theta_z < 113^{\circ}$ (see \tabref{event_samples}). Above a few GeV, the energy spectrum of the Limb is well modeled as a power law \cite{Ackermann:2014ula}. Therefore we expect the power-law approximation of background spectra to be better than in the GP. Below a few GeV, modeling the Limb spectrum becomes complicated due to the geomagnetic cutoff, so we only fit for line-like features in the Limb with \Egamma $>$ 5 GeV.

We select a $20^{\circ}$ ROI around the Vela pulsar using the `Celestial' data selections outlined in \tabref{event_samples}. In our Vela analysis, we use pulsar phases calculated with the \emph{Tempo2} package\footnote{\url{http://www.atnf.csiro.au/research/pulsar/tempo2/}}~\cite{REF:2006:TEMPO2}
and the standard ephemeris\footnote{\url{http://fermi.gsfc.nasa.gov/ssc/data/access/lat/ephems/}}.  For our line fits, we select the on-pulse data (\gammaRays\ with phases in the ranges $[0.1,0.3]\cup[0.5,0.6]$) and model the background in energy bin $i$ as a sum of the off-pulse (\gammaRays\ with phases in the range $[0.7,1.0]$) spectrum and an exponential cutoff model

\begin{linenomath}
\begin{equation}\label{eq:VelaModel}
   C_{\rm{bkg,Vela}}= \alpha\left(\frac{\Ereco}{\Epivot}\right)^{-\GamBkg}\exp[-(E/E_{\rm c})]\mathcal{E}(E),
\end{equation}
\end{linenomath}

\noindent where $\alpha$ is the normalization factor, $\Epivot$ is set to 100~MeV.  We fix $E_{\rm c} = 3$~GeV (which is slightly different from those cited in Ref. \cite{REF:2010.VelaII}) and let $\Gamma$ float free in the fit. Since the Vela spectrum cuts off steeply above a few GeV, we only fit for lines up to \Egamma\ = 10 GeV, making this region complementary to the higher-energy studies performed using the Limb data. The observed fractional signal values in the GP, Vela, and the Limb are shown in \figref{control_fit_examples}.

\onepanel{ht}{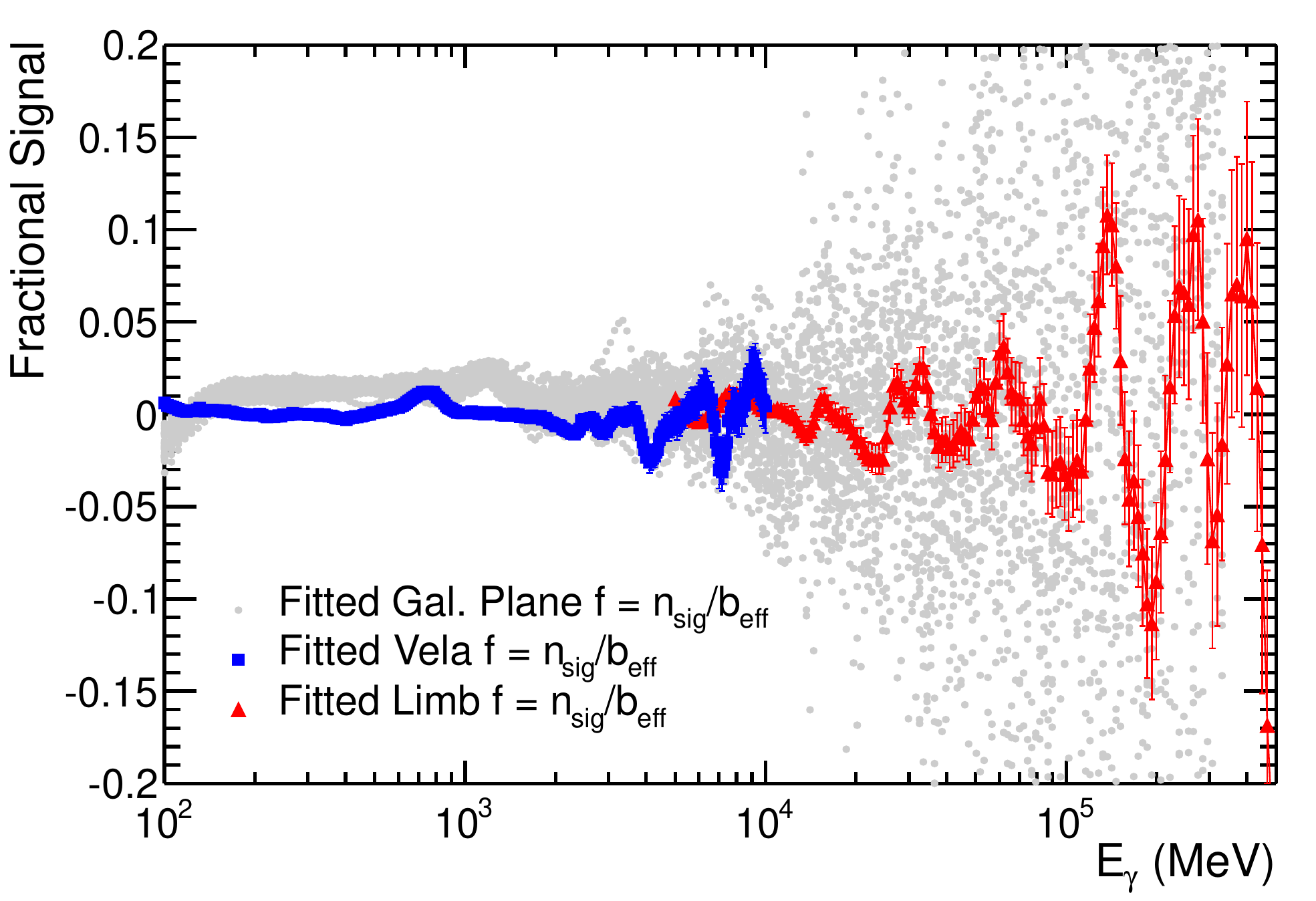}{
  \caption{(right) Fractional signals observed in the the Galactic plane (GP) scan (grey dots), from the Vela pulsar (blue squares), and from the Earth's Limb (red triangles). The error bars for the Vela and Limb points are $\fstat = 1 / \sqrt{\beff}$.}
  \label{fig:control_fit_examples}
}

Note that the features observed in the Limb, Vela, and the GP do not have the same energy dependence. This is because the inaccuracies of the background model are different in each control region and these errors induced different line-like features, especially from the modeling of the background spectra. For example, a slight increase of $f$ values is observed in the GP around 1.5 GeV, and a similar increase is observed in the Vela scans, but closer to 800 MeV instead. 
Both background models approximate the energy-dependent exposure similarly, but the flux models differ since the intrinsic \gammaRayHyph\ emission in the GP does not cut off like the Vela emission does. 
Therefore, the deficiencies in the total background spectral modeling induce $f$ values with different energy dependencies.  However, we note that the magnitude of the induced fractional signals is similar across all control regions. This motivated our choice of modeling \fsyst\ with a general Gaussian envelope centered on zero instead of using a more sophisticated energy-dependent model.

\section{95\% CONFIDENCE LEVEL LIMITS}\label{app:results}
We present the 95\% CL flux upper limits derived for each of our
5 ROIs in Tabs.~\ref{tab:allLimits0}--\ref{tab:allLimits2}.
We also give the annihilation cross section
upper limits for the DM profiles in each ROI where sensitivity to that DM model has been optimized: R3 (contracted
NFW profile),  R16 (Einasto profile), R41 (NFW profile) and R90
(Isothermal profile), and the decay lifetime lower limit for R180 (NFW profile).

\begin{table}[ht]
\caption{\label{tab:allLimits0}95\% confidence level limits from all ROIs for fit energies from 0.214--5.22 GeV.  The first column for each ROI is the ($\Phi_{\gamma\gamma}$) upper limit in $10^{-8}$ cm$^{2}$s$^{-1}$.  The second column for each is the upper limit on $\langle\sigma v\rangle_{\gamma\gamma}$ in $10^{-30}$ cm$^{3}$s$^{-1}$ for the DM profile for which that ROI is optimal.  For R180, we state lower limit on $\tau_{\gamma\nu}$(NFW) in $10^{28}$ s.  Note that for $\tau_{\gamma\nu}$, the energy is $m_{\chi}/2$.}
\begin{ruledtabular}
\begin{tabular}{ccccccccccc}
& \multicolumn{2}{c}{R3} & \multicolumn{2}{c}{R16} & \multicolumn{2}{c}{R41} & \multicolumn{2}{c}{R90} & \multicolumn{2}{c}{R180} \\
Energy &  $\Phi_{\gamma\gamma}$ & $\langle\sigma v \rangle_{\gamma\gamma}$ & $\Phi_{\gamma\gamma}$ & $\langle\sigma v \rangle_{\gamma\gamma}$ & $\Phi_{\gamma\gamma}$ & $\langle\sigma v \rangle_{\gamma\gamma}$ & $\Phi_{\gamma\gamma}$ & $\langle\sigma v \rangle_{\gamma\gamma}$ & $\Phi_{\gamma\gamma}$ & $\tau_{\gamma\nu}$ \\
(GeV) &  & NFWc & & Ein &  & NFW &  & Iso &  & NFW \\
0.214 & 18.4 & 0.766 & 45.5 & 2.79 & 135 & 8.52 & 315 & 25.8 & 587 & 0.788 \\
0.234 & 15.7 & 0.778 & 38.0 & 2.79 & 117 & 8.78 & 280 & 27.4 & 530 & 0.798 \\
0.255 & 14.8 & 0.873 & 34.0 & 2.97 & 102 & 9.14 & 248 & 28.8 & 477 & 0.813 \\
0.278 & 14.8 & 1.03 & 32.4 & 3.35 & 96.4 & 10.2 & 228 & 31.5 & 440 & 0.810 \\
0.303 & 13.2 & 1.10 & 30.9 & 3.79 & 92.0 & 11.6 & 210 & 34.4 & 399 & 0.820 \\
0.329 & 12.8 & 1.26 & 28.9 & 4.20 & 85.5 & 12.7 & 189 & 36.6 & 353 & 0.852 \\
0.358 & 11.4 & 1.32 & 26.3 & 4.51 & 82.4 & 14.5 & 168 & 38.4 & 310 & 0.893 \\
0.388 & 10.1 & 1.37 & 21.5 & 4.33 & 65.5 & 13.5 & 141 & 37.9 & 264 & 0.965 \\
0.421 & 9.16 & 1.47 & 18.8 & 4.47 & 56.5 & 13.7 & 121 & 38.2 & 228 & 1.03 \\
0.456 & 8.59 & 1.61 & 16.7 & 4.66 & 48.9 & 13.9 & 105 & 39.2 & 199 & 1.09 \\
0.493 & 7.42 & 1.63 & 14.7 & 4.78 & 42.2 & 14.1 & 89.2 & 38.8 & 171 & 1.18 \\
0.533 & 6.45 & 1.66 & 13.5 & 5.13 & 37.5 & 14.6 & 78.8 & 40.0 & 149 & 1.25 \\
0.576 & 5.71 & 1.71 & 11.8 & 5.23 & 33.8 & 15.3 & 70.7 & 41.8 & 134 & 1.28 \\
0.620 & 4.66 & 1.62 & 10.4 & 5.35 & 30.6 & 16.2 & 63.7 & 43.8 & 121 & 1.32 \\
0.668 & 3.83 & 1.54 & 9.84 & 5.88 & 29.5 & 18.1 & 62.0 & 49.3 & 115 & 1.29 \\
0.718 & 3.26 & 1.52 & 8.78 & 6.06 & 26.8 & 19.0 & 57.4 & 52.8 & 107 & 1.29 \\
0.770 & 3.30 & 1.77 & 7.80 & 6.20 & 23.1 & 18.8 & 49.3 & 52.3 & 76.3 & 1.69 \\
0.826 & 2.85 & 1.76 & 6.74 & 6.16 & 19.6 & 18.3 & 39.8 & 48.4 & 73.4 & 1.64 \\
0.885 & 2.26 & 1.60 & 6.36 & 6.66 & 17.6 & 18.9 & 33.6 & 47.0 & 61.3 & 1.83 \\
0.947 & 2.52 & 2.05 & 7.32 & 8.79 & 19.3 & 23.7 & 37.1 & 59.5 & 68.7 & 1.52 \\
1.01 & 2.25 & 2.09 & 8.68 & 11.9 & 22.6 & 31.8 & 44.8 & 82.1 & 81.9 & 1.19 \\
1.08 & 2.45 & 2.60 & 9.29 & 14.6 & 24.6 & 39.6 & 50.5 & 106 & 92.1 & 0.993 \\
1.16 & 2.22 & 2.69 & 8.68 & 15.6 & 23.3 & 42.8 & 49.0 & 117 & 89.5 & 0.956 \\
1.24 & 1.82 & 2.51 & 7.55 & 15.4 & 20.8 & 43.7 & 43.8 & 119 & 79.3 & 1.01 \\
1.32 & 1.47 & 2.31 & 6.39 & 14.9 & 18.0 & 42.9 & 37.0 & 115 & 67.1 & 1.12 \\
1.41 & 1.29 & 2.30 & 5.12 & 13.5 & 14.6 & 39.6 & 29.7 & 105 & 53.8 & 1.31 \\
1.50 & 1.21 & 2.45 & 4.03 & 12.1 & 11.5 & 35.4 & 23.1 & 92.3 & 41.3 & 1.60 \\
1.60 & 0.936 & 2.16 & 3.17 & 10.8 & 8.66 & 30.3 & 17.6 & 79.8 & 31.9 & 1.95 \\
1.70 & 0.815 & 2.13 & 2.92 & 11.3 & 7.12 & 28.3 & 14.2 & 73.2 & 25.6 & 2.27 \\
1.81 & 0.596 & 1.77 & 2.20 & 9.68 & 5.66 & 25.5 & 11.1 & 65.1 & 20.5 & 2.67 \\
1.93 & 0.430 & 1.45 & 1.84 & 9.18 & 4.70 & 24.1 & 9.98 & 66.4 & 17.9 & 2.87 \\
2.06 & 0.428 & 1.64 & 1.93 & 11.0 & 4.68 & 27.2 & 9.70 & 73.3 & 17.0 & 2.84 \\
2.19 & 0.461 & 2.00 & 1.87 & 12.1 & 4.63 & 30.5 & 8.91 & 76.4 & 15.7 & 2.88 \\
2.33 & 0.506 & 2.50 & 1.75 & 12.8 & 4.23 & 31.6 & 7.78 & 75.8 & 13.7 & 3.09 \\
2.49 & 0.365 & 2.04 & 1.35 & 11.2 & 3.47 & 29.4 & 6.88 & 76.0 & 12.0 & 3.32 \\
2.65 & 0.234 & 1.49 & 1.07 & 10.1 & 2.87 & 27.6 & 6.19 & 77.5 & 10.7 & 3.51 \\
2.82 & 0.239 & 1.72 & 0.844 & 8.99 & 2.48 & 27.1 & 5.61 & 79.8 & 9.70 & 3.62 \\
3.00 & 0.282 & 2.31 & 0.738 & 8.92 & 2.31 & 28.6 & 5.26 & 84.8 & 8.63 & 3.82 \\
3.20 & 0.222 & 2.05 & 0.780 & 10.7 & 2.22 & 31.1 & 4.28 & 78.1 & 7.30 & 4.25 \\
3.40 & 0.289 & 3.03 & 0.915 & 14.2 & 2.07 & 32.9 & 3.85 & 79.5 & 6.47 & 4.50 \\
3.62 & 0.326 & 3.86 & 1.02 & 17.9 & 2.11 & 38.0 & 3.22 & 75.3 & 5.76 & 4.75 \\
3.85 & 0.321 & 4.31 & 0.925 & 18.4 & 1.95 & 39.7 & 2.83 & 74.9 & 4.91 & 5.24 \\
4.09 & 0.271 & 4.11 & 0.764 & 17.2 & 1.52 & 35.0 & 2.43 & 72.7 & 4.01 & 6.03 \\
4.35 & 0.175 & 3.00 & 0.715 & 18.1 & 1.25 & 32.5 & 2.22 & 75.1 & 3.66 & 6.21 \\
4.63 & 0.180 & 3.48 & 0.561 & 16.0 & 1.11 & 32.6 & 2.00 & 76.4 & 3.15 & 6.80 \\
4.91 & 0.157 & 3.43 & 0.445 & 14.4 & 0.964 & 31.9 & 1.68 & 72.5 & 2.84 & 7.10 \\
5.22 & 0.115 & 2.82 & 0.263 & 9.59 & 0.656 & 24.5 & 1.33 & 64.7 & 2.35 & 8.10 \\

\end{tabular}
\end{ruledtabular}
\end{table}

\begin{table}[ht]
\caption{\label{tab:allLimits1}95\% confidence level limits from all ROIs for fit energies from 5.54--43.8 GeV.  The first column for each ROI is the ($\Phi_{\gamma\gamma}$) upper limit in $10^{-10}$ cm$^{2}$s$^{-1}$.  The second column for each is the upper limit on $\langle\sigma v\rangle_{\gamma\gamma}$ in $10^{-29}$ cm$^{3}$s$^{-1}$ for the DM profile for which that ROI is optimal.  For R180, we state lower limit on $\tau_{\gamma\nu}$(NFW) in $10^{28}$ s.  Note that for $\tau_{\gamma\nu}$, the energy is $m_{\chi}/2$.}
\begin{ruledtabular}
\begin{tabular}{ccccccccccc}
& \multicolumn{2}{c}{R3} & \multicolumn{2}{c}{R16} & \multicolumn{2}{c}{R41} & \multicolumn{2}{c}{R90} & \multicolumn{2}{c}{R180} \\
Energy &  $\Phi_{\gamma\gamma}$ & $\langle\sigma v \rangle_{\gamma\gamma}$ & $\Phi_{\gamma\gamma}$ & $\langle\sigma v \rangle_{\gamma\gamma}$ & $\Phi_{\gamma\gamma}$ & $\langle\sigma v \rangle_{\gamma\gamma}$ & $\Phi_{\gamma\gamma}$ & $\langle\sigma v \rangle_{\gamma\gamma}$ & $\Phi_{\gamma\gamma}$ & $\tau_{\gamma\nu}$ \\
(GeV) &  & NFWc & & Ein &  & NFW &  & Iso &  & NFW \\
5.54 & 4.58 & 0.127 & 21.6 & 0.888 & 64.1 & 2.69 & 110 & 6.04 & 196 & 9.13 \\
5.87 & 4.62 & 0.144 & 20.6 & 0.950 & 63.7 & 3.01 & 102 & 6.27 & 187 & 9.03 \\
6.23 & 5.65 & 0.198 & 17.0 & 0.881 & 61.0 & 3.25 & 115 & 7.96 & 220 & 7.23 \\
6.60 & 7.84 & 0.309 & 22.5 & 1.31 & 75.0 & 4.48 & 144 & 11.2 & 263 & 5.70 \\
6.99 & 4.74 & 0.209 & 26.0 & 1.70 & 74.2 & 4.98 & 160 & 13.9 & 271 & 5.23 \\
7.40 & 5.43 & 0.269 & 33.4 & 2.45 & 76.2 & 5.73 & 152 & 14.8 & 259 & 5.16 \\
7.83 & 9.21 & 0.510 & 25.3 & 2.08 & 55.0 & 4.62 & 119 & 13.1 & 211 & 6.01 \\
8.28 & 6.14 & 0.381 & 18.6 & 1.71 & 39.8 & 3.74 & 100 & 12.3 & 177 & 6.76 \\
8.76 & 4.63 & 0.321 & 22.1 & 2.27 & 37.5 & 3.95 & 91.0 & 12.5 & 153 & 7.37 \\
9.26 & 3.62 & 0.281 & 12.7 & 1.46 & 28.8 & 3.39 & 74.2 & 11.4 & 131 & 8.19 \\
9.79 & 3.42 & 0.297 & 7.59 & 0.974 & 33.0 & 4.35 & 61.7 & 10.6 & 115 & 8.81 \\
10.4 & 5.61 & 0.543 & 8.14 & 1.17 & 28.9 & 4.25 & 59.6 & 11.4 & 106 & 8.99 \\
10.9 & 6.51 & 0.705 & 12.4 & 1.98 & 35.3 & 5.80 & 78.9 & 16.9 & 121 & 7.48 \\
11.6 & 6.41 & 0.775 & 16.0 & 2.87 & 36.4 & 6.67 & 73.9 & 17.6 & 117 & 7.32 \\
12.2 & 7.26 & 0.980 & 7.72 & 1.54 & 21.3 & 4.35 & 59.9 & 16.0 & 111 & 7.28 \\
12.9 & 5.15 & 0.775 & 7.83 & 1.75 & 23.2 & 5.30 & 56.7 & 16.9 & 111 & 6.91 \\
13.6 & 2.97 & 0.499 & 8.42 & 2.09 & 21.9 & 5.57 & 41.3 & 13.7 & 84.1 & 8.64 \\
14.4 & 2.86 & 0.536 & 5.91 & 1.64 & 15.3 & 4.34 & 27.4 & 10.2 & 60.9 & 11.3 \\
15.2 & 3.00 & 0.626 & 6.52 & 2.02 & 15.8 & 5.01 & 30.1 & 12.4 & 61.4 & 10.6 \\
16.1 & 2.76 & 0.645 & 6.66 & 2.30 & 17.1 & 6.06 & 40.6 & 18.7 & 68.4 & 9.02 \\
17.0 & 4.07 & 1.06 & 8.18 & 3.15 & 23.0 & 9.08 & 39.8 & 20.4 & 68.4 & 8.54 \\
17.9 & 2.50 & 0.726 & 10.2 & 4.37 & 20.8 & 9.19 & 42.5 & 24.4 & 62.6 & 8.83 \\
18.9 & 3.40 & 1.10 & 5.79 & 2.78 & 13.8 & 6.79 & 34.9 & 22.4 & 50.2 & 10.4 \\
20.0 & 5.54 & 2.00 & 3.65 & 1.96 & 11.3 & 6.21 & 19.6 & 14.0 & 32.5 & 15.3 \\
21.1 & 4.14 & 1.67 & 6.56 & 3.92 & 11.2 & 6.84 & 12.2 & 9.74 & 22.3 & 21.0 \\
22.3 & 2.10 & 0.946 & 3.66 & 2.44 & 5.57 & 3.81 & 9.01 & 8.02 & 18.7 & 23.7 \\
23.6 & 1.90 & 0.957 & 3.74 & 2.78 & 6.43 & 4.91 & 9.89 & 9.83 & 17.7 & 23.7 \\
24.9 & 1.47 & 0.829 & 2.97 & 2.47 & 8.70 & 7.43 & 14.8 & 16.4 & 24.2 & 16.4 \\
26.4 & 1.41 & 0.888 & 3.78 & 3.51 & 11.4 & 10.9 & 19.7 & 24.5 & 38.1 & 9.86 \\
27.9 & 1.77 & 1.24 & 4.56 & 4.74 & 14.4 & 15.3 & 29.8 & 41.3 & 46.3 & 7.67 \\
29.5 & 1.22 & 0.958 & 7.05 & 8.19 & 15.7 & 18.7 & 26.4 & 40.9 & 44.9 & 7.50 \\
31.2 & 0.753 & 0.661 & 4.37 & 5.68 & 9.49 & 12.6 & 20.5 & 35.5 & 37.5 & 8.47 \\
33.0 & 0.708 & 0.695 & 3.28 & 4.77 & 4.12 & 6.14 & 13.7 & 26.6 & 31.2 & 9.62 \\
34.9 & 1.29 & 1.42 & 4.17 & 6.79 & 5.95 & 9.93 & 14.9 & 32.4 & 25.4 & 11.2 \\
36.9 & 2.51 & 3.08 & 4.71 & 8.58 & 10.6 & 19.9 & 23.2 & 56.4 & 28.8 & 9.34 \\
39.0 & 2.81 & 3.87 & 3.18 & 6.48 & 9.47 & 19.8 & 18.4 & 50.1 & 25.4 & 10.0 \\
41.3 & 2.45 & 3.79 & 3.07 & 7.01 & 7.00 & 16.4 & 14.0 & 42.8 & 23.1 & 10.4 \\
43.8 & 3.70 & 6.41 & 4.71 & 12.1 & 7.13 & 18.8 & 12.7 & 43.6 & 20.7 & 11.0 \\
\end{tabular}
\end{ruledtabular}
\end{table}

\begin{table}[ht]
\caption{\label{tab:allLimits2}95\% confidence level limits from all ROIs for fit energies from 46.4--462 GeV.  The first column for each ROI is the ($\Phi_{\gamma\gamma}$) upper limit in $10^{-10}$ cm$^{2}$s$^{-1}$.  The second column for each is the upper limit on $\langle\sigma v\rangle_{\gamma\gamma}$ in $10^{-29}$ cm$^{3}$s$^{-1}$ for the DM profile for which that ROI is optimal.  For R180, we state lower limit on $\tau_{\gamma\nu}$(NFW) in $10^{28}$ s.  Note that for $\tau_{\gamma\nu}$, the energy is $m_{\chi}/2$.}
\begin{ruledtabular}
\begin{tabular}{ccccccccccc}
& \multicolumn{2}{c}{R3} & \multicolumn{2}{c}{R16} & \multicolumn{2}{c}{R41} & \multicolumn{2}{c}{R90} & \multicolumn{2}{c}{R180} \\
Energy &  $\Phi_{\gamma\gamma}$ & $\langle\sigma v \rangle_{\gamma\gamma}$ & $\Phi_{\gamma\gamma}$ & $\langle\sigma v \rangle_{\gamma\gamma}$ & $\Phi_{\gamma\gamma}$ & $\langle\sigma v \rangle_{\gamma\gamma}$ & $\Phi_{\gamma\gamma}$ & $\langle\sigma v \rangle_{\gamma\gamma}$ & $\Phi_{\gamma\gamma}$ & $\tau_{\gamma\nu}$ \\
(GeV) &  & NFWc & & Ein &  & NFW &  & Iso &  & NFW \\
46.4 & 3.37 & 6.54 & 5.66 & 16.3 & 7.64 & 22.5 & 11.6 & 44.7 & 16.4 & 13.0 \\
49.1 & 2.58 & 5.62 & 6.40 & 20.7 & 5.66 & 18.7 & 6.18 & 26.6 & 10.9 & 18.4 \\
52.1 & 1.27 & 3.12 & 4.56 & 16.6 & 3.07 & 11.4 & 4.55 & 22.0 & 9.86 & 19.3 \\
55.2 & 1.23 & 3.38 & 3.96 & 16.2 & 3.88 & 16.2 & 4.73 & 25.7 & 10.2 & 17.7 \\
58.6 & 2.30 & 7.13 & 4.85 & 22.3 & 5.87 & 27.7 & 7.28 & 44.6 & 10.0 & 16.9 \\
62.2 & 1.99 & 6.95 & 3.32 & 17.2 & 2.85 & 15.1 & 4.38 & 30.2 & 5.77 & 27.6 \\
66.0 & 1.17 & 4.59 & 1.82 & 10.6 & 2.11 & 12.6 & 5.03 & 39.1 & 5.38 & 27.9 \\
70.1 & 1.17 & 5.18 & 1.90 & 12.5 & 4.49 & 30.2 & 5.92 & 51.9 & 5.61 & 25.2 \\
74.5 & 1.11 & 5.56 & 3.63 & 26.9 & 6.61 & 50.3 & 4.81 & 47.6 & 9.46 & 14.1 \\
79.2 & 0.543 & 3.08 & 1.48 & 12.4 & 3.31 & 28.5 & 4.27 & 47.8 & 8.32 & 15.0 \\
84.2 & 0.448 & 2.87 & 0.951 & 9.03 & 1.88 & 18.3 & 2.82 & 35.7 & 4.95 & 23.8 \\
89.6 & 0.396 & 2.87 & 0.947 & 10.2 & 2.36 & 26.0 & 4.20 & 60.2 & 6.77 & 16.3 \\
95.4 & 0.343 & 2.82 & 0.891 & 10.8 & 2.88 & 35.9 & 7.63 & 124 & 9.86 & 10.5 \\
102 & 0.619 & 5.77 & 2.29 & 31.6 & 4.25 & 60.0 & 6.57 & 121 & 9.60 & 10.2 \\
108 & 0.541 & 5.73 & 4.89 & 76.5 & 5.89 & 94.6 & 5.60 & 117 & 5.26 & 17.4 \\
115 & 1.26 & 15.2 & 4.92 & 87.5 & 7.45 & 136 & 6.11 & 145 & 5.47 & 15.7 \\
123 & 1.11 & 15.1 & 3.84 & 77.7 & 6.04 & 125 & 4.17 & 113 & 3.64 & 22.1 \\
131 & 0.693 & 10.8 & 3.11 & 71.7 & 4.49 & 106 & 5.09 & 156 & 5.11 & 14.8 \\
140 & 0.298 & 5.29 & 1.48 & 39.0 & 3.17 & 85.5 & 4.44 & 156 & 4.54 & 15.6 \\
150 & 0.523 & 10.6 & 0.765 & 22.9 & 1.54 & 47.2 & 4.31 & 172 & 6.22 & 10.6 \\
160 & 0.352 & 8.15 & 0.764 & 26.2 & 1.04 & 36.4 & 2.86 & 131 & 5.71 & 10.9 \\
171 & 0.492 & 13.0 & 1.11 & 43.4 & 1.56 & 62.5 & 3.12 & 163 & 4.88 & 11.9 \\
183 & 0.221 & 6.68 & 1.70 & 76.4 & 2.50 & 115 & 3.64 & 217 & 4.61 & 11.7 \\
196 & 0.385 & 13.3 & 2.22 & 114 & 2.94 & 155 & 2.57 & 176 & 3.29 & 15.4 \\
210 & 0.231 & 9.19 & 2.85 & 168 & 3.72 & 225 & 1.65 & 130 & 2.30 & 20.6 \\
225 & 0.286 & 13.0 & 1.59 & 107 & 4.86 & 337 & 2.33 & 210 & 5.29 & 8.34 \\
241 & 0.356 & 18.7 & 1.93 & 150 & 3.12 & 248 & 2.99 & 309 & 5.24 & 7.85 \\
259 & 0.169 & 10.2 & 0.867 & 77.5 & 0.813 & 74.6 & 1.78 & 213 & 2.52 & 15.2 \\
276 & 0.600 & 41.2 & 0.843 & 85.7 & 0.820 & 85.4 & 1.67 & 227 & 1.99 & 18.0 \\
294 & 0.529 & 41.4 & 1.32 & 153 & 1.51 & 179 & 3.34 & 516 & 2.24 & 15.0 \\
321 & 0.186 & 17.3 & 1.45 & 200 & 1.78 & 251 & 3.63 & 666 & 2.05 & 15.1 \\
345 & 0.140 & 15.0 & 1.17 & 186 & 1.13 & 184 & 2.15 & 456 & 1.95 & 14.7 \\
367 & 0.401 & 48.7 & 0.646 & 116 & 0.807 & 149 & 2.06 & 494 & 2.15 & 12.6 \\
396 & 0.366 & 51.7 & 0.613 & 128 & 1.04 & 223 & 1.39 & 387 & 1.86 & 13.5 \\
427 & 0.301 & 49.6 & 0.560 & 137 & 1.67 & 418 & 0.896 & 292 & 1.86 & 12.5 \\
462 & 0.204 & 39.4 & 0.487 & 139 & 1.16 & 341 & 0.768 & 293 & 1.23 & 17.5 \\

\end{tabular}
\end{ruledtabular}
\end{table}

\bibliography{P8Line}

\begin{thebibliography}{38}%
\makeatletter
\providecommand \@ifxundefined [1]{%
 \@ifx{#1\undefined}
}%
\providecommand \@ifnum [1]{%
 \ifnum #1\expandafter \@firstoftwo
 \else \expandafter \@secondoftwo
 \fi
}%
\providecommand \@ifx [1]{%
 \ifx #1\expandafter \@firstoftwo
 \else \expandafter \@secondoftwo
 \fi
}%
\providecommand \natexlab [1]{#1}%
\providecommand \enquote  [1]{``#1''}%
\providecommand \bibnamefont  [1]{#1}%
\providecommand \bibfnamefont [1]{#1}%
\providecommand \citenamefont [1]{#1}%
\providecommand \href@noop [0]{\@secondoftwo}%
\providecommand \href [0]{\begingroup \@sanitize@url \@href}%
\providecommand \@href[1]{\@@startlink{#1}\@@href}%
\providecommand \@@href[1]{\endgroup#1\@@endlink}%
\providecommand \@sanitize@url [0]{\catcode `\\12\catcode `\$12\catcode
  `\&12\catcode `\#12\catcode `\^12\catcode `\_12\catcode `\%12\relax}%
\providecommand \@@startlink[1]{}%
\providecommand \@@endlink[0]{}%
\providecommand \url  [0]{\begingroup\@sanitize@url \@url }%
\providecommand \@url [1]{\endgroup\@href {#1}{\urlprefix }}%
\providecommand \urlprefix  [0]{URL }%
\providecommand \Eprint [0]{\href }%
\providecommand \doibase [0]{http://dx.doi.org/}%
\providecommand \selectlanguage [0]{\@gobble}%
\providecommand \bibinfo  [0]{\@secondoftwo}%
\providecommand \bibfield  [0]{\@secondoftwo}%
\providecommand \translation [1]{[#1]}%
\providecommand \BibitemOpen [0]{}%
\providecommand \bibitemStop [0]{}%
\providecommand \bibitemNoStop [0]{.\EOS\space}%
\providecommand \EOS [0]{\spacefactor3000\relax}%
\providecommand \BibitemShut  [1]{\csname bibitem#1\endcsname}%
\let\auto@bib@innerbib\@empty
\bibitem [{\citenamefont {Ade}\ \emph {et~al.}(2013)\citenamefont {Ade} \emph
  {et~al.}}]{Ade:2013lta}%
  \BibitemOpen
  \bibfield  {author} {\bibinfo {author} {\bibfnamefont {P.}~\bibnamefont
  {Ade}} \emph {et~al.} (\bibinfo {collaboration} {Planck Collaboration}),\
  }\href@noop {} {\  (\bibinfo {year} {2013})},\ \Eprint
  {http://arxiv.org/abs/1303.5076} {arXiv:1303.5076 [astro-ph.CO]} \BibitemShut
  {NoStop}%
\bibitem [{\citenamefont {Sofue}\ and\ \citenamefont
  {Rubin}(2001)}]{Sofue:2000jx}%
  \BibitemOpen
  \bibfield  {author} {\bibinfo {author} {\bibfnamefont {Y.}~\bibnamefont
  {Sofue}}\ and\ \bibinfo {author} {\bibfnamefont {V.}~\bibnamefont {Rubin}},\
  }\href {\doibase 10.1146/annurev.astro.39.1.137} {\bibfield  {journal}
  {\bibinfo  {journal} {Ann.Rev.Astron.Astrophys.}\ }\textbf {\bibinfo {volume}
  {39}},\ \bibinfo {pages} {137} (\bibinfo {year} {2001})},\ \Eprint
  {http://arxiv.org/abs/astro-ph/0010594} {arXiv:astro-ph/0010594 [astro-ph]}
  \BibitemShut {NoStop}%
\bibitem [{\citenamefont {Clowe}\ \emph {et~al.}(2006)\citenamefont {Clowe}
  \emph {et~al.}}]{Clowe:2006eq}%
  \BibitemOpen
  \bibfield  {author} {\bibinfo {author} {\bibfnamefont {D.}~\bibnamefont
  {Clowe}} \emph {et~al.},\ }\href {\doibase 10.1086/508162} {\bibfield
  {journal} {\bibinfo  {journal} {Astrophys.J.}\ }\textbf {\bibinfo {volume}
  {648}},\ \bibinfo {pages} {L109} (\bibinfo {year} {2006})},\ \Eprint
  {http://arxiv.org/abs/astro-ph/0608407} {arXiv:astro-ph/0608407 [astro-ph]}
  \BibitemShut {NoStop}%
\bibitem [{\citenamefont {Bertone}\ \emph {et~al.}(2005)\citenamefont
  {Bertone}, \citenamefont {Hooper},\ and\ \citenamefont
  {Silk}}]{Bertone:2004pz}%
  \BibitemOpen
  \bibfield  {author} {\bibinfo {author} {\bibfnamefont {G.}~\bibnamefont
  {Bertone}}, \bibinfo {author} {\bibfnamefont {D.}~\bibnamefont {Hooper}}, \
  and\ \bibinfo {author} {\bibfnamefont {J.}~\bibnamefont {Silk}},\ }\href
  {\doibase 10.1016/j.physrep.2004.08.031} {\bibfield  {journal} {\bibinfo
  {journal} {Phys.Rept.}\ }\textbf {\bibinfo {volume} {405}},\ \bibinfo {pages}
  {279} (\bibinfo {year} {2005})},\ \Eprint
  {http://arxiv.org/abs/hep-ph/0404175} {arXiv:hep-ph/0404175 [hep-ph]}
  \BibitemShut {NoStop}%
\bibitem [{\citenamefont {Feng}(2010)}]{Feng:2010gw}%
  \BibitemOpen
  \bibfield  {author} {\bibinfo {author} {\bibfnamefont {J.~L.}\ \bibnamefont
  {Feng}},\ }\href {\doibase 10.1146/annurev-astro-082708-101659} {\bibfield
  {journal} {\bibinfo  {journal} {Ann.Rev.Astron.Astrophys.}\ }\textbf
  {\bibinfo {volume} {48}},\ \bibinfo {pages} {495} (\bibinfo {year} {2010})},\
  \Eprint {http://arxiv.org/abs/1003.0904} {arXiv:1003.0904 [astro-ph.CO]}
  \BibitemShut {NoStop}%
\bibitem [{\citenamefont {Bringmann}\ and\ \citenamefont
  {Weniger}(2012)}]{Bringmann:2012ez}%
  \BibitemOpen
  \bibfield  {author} {\bibinfo {author} {\bibfnamefont {T.}~\bibnamefont
  {Bringmann}}\ and\ \bibinfo {author} {\bibfnamefont {C.}~\bibnamefont
  {Weniger}},\ }\href {\doibase 10.1016/j.dark.2012.10.005} {\bibfield
  {journal} {\bibinfo  {journal} {Phys.Dark Univ.}\ }\textbf {\bibinfo {volume}
  {1}},\ \bibinfo {pages} {194} (\bibinfo {year} {2012})},\ \Eprint
  {http://arxiv.org/abs/1208.5481} {arXiv:1208.5481 [hep-ph]} \BibitemShut
  {NoStop}%
\bibitem [{\citenamefont {Ibarra}\ \emph {et~al.}(2013)\citenamefont {Ibarra},
  \citenamefont {Tran},\ and\ \citenamefont {Weniger}}]{Ibarra:2013cra}%
  \BibitemOpen
  \bibfield  {author} {\bibinfo {author} {\bibfnamefont {A.}~\bibnamefont
  {Ibarra}}, \bibinfo {author} {\bibfnamefont {D.}~\bibnamefont {Tran}}, \ and\
  \bibinfo {author} {\bibfnamefont {C.}~\bibnamefont {Weniger}},\ }\href
  {\doibase 10.1142/S0217751X13300408} {\bibfield  {journal} {\bibinfo
  {journal} {Int.J.Mod.Phys.}\ }\textbf {\bibinfo {volume} {A28}},\ \bibinfo
  {pages} {1330040} (\bibinfo {year} {2013})},\ \Eprint
  {http://arxiv.org/abs/1307.6434} {arXiv:1307.6434 [hep-ph]} \BibitemShut
  {NoStop}%
\bibitem [{\citenamefont {{Aharonian}}\ \emph {et~al.}(2012)\citenamefont
  {{Aharonian}}, \citenamefont {{Khangulyan}},\ and\ \citenamefont
  {{Malyshev}}}]{Aharonian:2012}%
  \BibitemOpen
  \bibfield  {author} {\bibinfo {author} {\bibfnamefont {F.}~\bibnamefont
  {{Aharonian}}}, \bibinfo {author} {\bibfnamefont {D.}~\bibnamefont
  {{Khangulyan}}}, \ and\ \bibinfo {author} {\bibfnamefont {D.}~\bibnamefont
  {{Malyshev}}},\ }\href {\doibase 10.1051/0004-6361/201220092} {\bibfield
  {journal} {\bibinfo  {journal} {\aap}\ }\textbf {\bibinfo {volume} {547}},\
  \bibinfo {eid} {A114} (\bibinfo {year} {2012})},\ \Eprint
  {http://arxiv.org/abs/1207.0458} {arXiv:1207.0458 [astro-ph.HE]} \BibitemShut
  {NoStop}%
\bibitem [{\citenamefont {Bergstrom}\ and\ \citenamefont
  {Ullio}(1997)}]{REF:Bergstrom:1997fh}%
  \BibitemOpen
  \bibfield  {author} {\bibinfo {author} {\bibfnamefont {L.}~\bibnamefont
  {Bergstrom}}\ and\ \bibinfo {author} {\bibfnamefont {P.}~\bibnamefont
  {Ullio}},\ }\href {\doibase 10.1016/S0550-3213(97)00530-0} {\bibfield
  {journal} {\bibinfo  {journal} {Nucl.Phys.}\ }\textbf {\bibinfo {volume}
  {B504}},\ \bibinfo {pages} {27} (\bibinfo {year} {1997})},\ \Eprint
  {http://arxiv.org/abs/hep-ph/9706232} {arXiv:hep-ph/9706232 [hep-ph]}
  \BibitemShut {NoStop}%
\bibitem [{\citenamefont {Matsumoto}\ \emph {et~al.}(2005)\citenamefont
  {Matsumoto}, \citenamefont {Sato},\ and\ \citenamefont
  {Sato}}]{REF:Matsumoto:2005ui}%
  \BibitemOpen
  \bibfield  {author} {\bibinfo {author} {\bibfnamefont {S.}~\bibnamefont
  {Matsumoto}}, \bibinfo {author} {\bibfnamefont {J.}~\bibnamefont {Sato}}, \
  and\ \bibinfo {author} {\bibfnamefont {Y.}~\bibnamefont {Sato}},\ }\href@noop
  {} {\  (\bibinfo {year} {2005})},\ \Eprint
  {http://arxiv.org/abs/hep-ph/0505160} {arXiv:hep-ph/0505160 [hep-ph]}
  \BibitemShut {NoStop}%
\bibitem [{\citenamefont {Ferrer}\ \emph {et~al.}(2006)\citenamefont {Ferrer},
  \citenamefont {Krauss},\ and\ \citenamefont {Profumo}}]{REF:Ferrer:2006hy}%
  \BibitemOpen
  \bibfield  {author} {\bibinfo {author} {\bibfnamefont {F.}~\bibnamefont
  {Ferrer}}, \bibinfo {author} {\bibfnamefont {L.~M.}\ \bibnamefont {Krauss}},
  \ and\ \bibinfo {author} {\bibfnamefont {S.}~\bibnamefont {Profumo}},\ }\href
  {\doibase 10.1103/PhysRevD.74.115007} {\bibfield  {journal} {\bibinfo
  {journal} {Phys.Rev.}\ }\textbf {\bibinfo {volume} {D74}},\ \bibinfo {pages}
  {115007} (\bibinfo {year} {2006})},\ \Eprint
  {http://arxiv.org/abs/hep-ph/0609257} {arXiv:hep-ph/0609257 [hep-ph]}
  \BibitemShut {NoStop}%
\bibitem [{\citenamefont {Gustafsson}\ \emph {et~al.}(2007)\citenamefont
  {Gustafsson} \emph {et~al.}}]{REF:Gustafsson:2007pc}%
  \BibitemOpen
  \bibfield  {author} {\bibinfo {author} {\bibfnamefont {M.}~\bibnamefont
  {Gustafsson}} \emph {et~al.},\ }\href {\doibase
  10.1103/PhysRevLett.99.041301} {\bibfield  {journal} {\bibinfo  {journal}
  {Phys.Rev.Lett.}\ }\textbf {\bibinfo {volume} {99}},\ \bibinfo {pages}
  {041301} (\bibinfo {year} {2007})},\ \Eprint
  {http://arxiv.org/abs/astro-ph/0703512} {arXiv:astro-ph/0703512 [ASTRO-PH]}
  \BibitemShut {NoStop}%
\bibitem [{\citenamefont {Profumo}(2008)}]{REF:Profumo:2008yg}%
  \BibitemOpen
  \bibfield  {author} {\bibinfo {author} {\bibfnamefont {S.}~\bibnamefont
  {Profumo}},\ }\href {\doibase 10.1103/PhysRevD.78.023507} {\bibfield
  {journal} {\bibinfo  {journal} {Phys.Rev.}\ }\textbf {\bibinfo {volume}
  {D78}},\ \bibinfo {pages} {023507} (\bibinfo {year} {2008})},\ \Eprint
  {http://arxiv.org/abs/0806.2150} {arXiv:0806.2150 [hep-ph]} \BibitemShut
  {NoStop}%
\bibitem [{\citenamefont {Steigman}\ \emph {et~al.}(2012)\citenamefont
  {Steigman}, \citenamefont {Dasgupta},\ and\ \citenamefont
  {Beacom}}]{Steigman:2012nb}%
  \BibitemOpen
  \bibfield  {author} {\bibinfo {author} {\bibfnamefont {G.}~\bibnamefont
  {Steigman}}, \bibinfo {author} {\bibfnamefont {B.}~\bibnamefont {Dasgupta}},
  \ and\ \bibinfo {author} {\bibfnamefont {J.~F.}\ \bibnamefont {Beacom}},\
  }\href {\doibase 10.1103/PhysRevD.86.023506} {\bibfield  {journal} {\bibinfo
  {journal} {Phys.Rev.}\ }\textbf {\bibinfo {volume} {D86}},\ \bibinfo {pages}
  {023506} (\bibinfo {year} {2012})},\ \Eprint {http://arxiv.org/abs/1204.3622}
  {arXiv:1204.3622 [hep-ph]} \BibitemShut {NoStop}%
\bibitem [{\citenamefont {{Atwood}}\ \emph {et~al.}(2009)\citenamefont
  {{Atwood}} \emph {et~al.}}]{REF:2009.LATPaper}%
  \BibitemOpen
  \bibfield  {author} {\bibinfo {author} {\bibfnamefont {W.~B.}\ \bibnamefont
  {{Atwood}}} \emph {et~al.} (\bibinfo {collaboration} {\fermi-LAT
  Collaboration}),\ }\href {\doibase 10.1088/0004-637X/697/2/1071} {\bibfield
  {journal} {\bibinfo  {journal} {\apj}\ }\textbf {\bibinfo {volume} {697}},\
  \bibinfo {pages} {1071} (\bibinfo {year} {2009})},\ \Eprint
  {http://arxiv.org/abs/0902.1089} {arXiv:0902.1089 [astro-ph.IM]} \BibitemShut
  {NoStop}%
\bibitem [{\citenamefont {Ibarra}\ and\ \citenamefont
  {Tran}(2008)}]{REF:Ibarra:2007wg}%
  \BibitemOpen
  \bibfield  {author} {\bibinfo {author} {\bibfnamefont {A.}~\bibnamefont
  {Ibarra}}\ and\ \bibinfo {author} {\bibfnamefont {D.}~\bibnamefont {Tran}},\
  }\href {\doibase 10.1103/PhysRevLett.100.061301} {\bibfield  {journal}
  {\bibinfo  {journal} {Phys.Rev.Lett.}\ }\textbf {\bibinfo {volume} {100}},\
  \bibinfo {pages} {061301} (\bibinfo {year} {2008})},\ \Eprint
  {http://arxiv.org/abs/0709.4593} {arXiv:0709.4593 [astro-ph]} \BibitemShut
  {NoStop}%
\bibitem [{\citenamefont {{Abdo}}\ \emph {et~al.}(2010)\citenamefont {{Abdo}}
  \emph {et~al.}}]{REF:2010.LineSearch}%
  \BibitemOpen
  \bibfield  {author} {\bibinfo {author} {\bibfnamefont {A.~A.}\ \bibnamefont
  {{Abdo}}} \emph {et~al.} (\bibinfo {collaboration} {\fermi-LAT
  Collaboration}),\ }\href {\doibase 10.1103/PhysRevLett.104.091302} {\bibfield
   {journal} {\bibinfo  {journal} {Physical Review Letters}\ }\textbf {\bibinfo
  {volume} {104}},\ \bibinfo {pages} {091302} (\bibinfo {year} {2010})},\
  \Eprint {http://arxiv.org/abs/1001.4836} {arXiv:1001.4836 [astro-ph.HE]}
  \BibitemShut {NoStop}%
\bibitem [{\citenamefont {{Ackermann}}\ \emph
  {et~al.}(2012{\natexlab{a}})\citenamefont {{Ackermann}} \emph
  {et~al.}}]{REF:2012.LineSearch}%
  \BibitemOpen
  \bibfield  {author} {\bibinfo {author} {\bibfnamefont {M.}~\bibnamefont
  {{Ackermann}}} \emph {et~al.} (\bibinfo {collaboration} {\fermi-LAT
  Collaboration}),\ }\href {\doibase 10.1103/PhysRevD.86.022002} {\bibfield
  {journal} {\bibinfo  {journal} {\prd}\ }\textbf {\bibinfo {volume} {86}},\
  \bibinfo {eid} {022002} (\bibinfo {year} {2012}{\natexlab{a}})},\ \Eprint
  {http://arxiv.org/abs/1205.2739} {arXiv:1205.2739 [astro-ph.HE]} \BibitemShut
  {NoStop}%
\bibitem [{\citenamefont {Ackermann}\ \emph {et~al.}(2013)\citenamefont
  {Ackermann} \emph {et~al.}}]{REF:P7Line}%
  \BibitemOpen
  \bibfield  {author} {\bibinfo {author} {\bibfnamefont {M.}~\bibnamefont
  {Ackermann}} \emph {et~al.} (\bibinfo {collaboration} {\fermi-LAT
  Collaboration}),\ }\href {\doibase 10.1103/PhysRevD.88.082002} {\bibfield
  {journal} {\bibinfo  {journal} {Phys.Rev.}\ }\textbf {\bibinfo {volume}
  {D88}},\ \bibinfo {pages} {082002} (\bibinfo {year} {2013})},\ \Eprint
  {http://arxiv.org/abs/1305.5597} {arXiv:1305.5597 [astro-ph.HE]} \BibitemShut
  {NoStop}%
\bibitem [{\citenamefont {Bringmann}\ \emph {et~al.}(2012)\citenamefont
  {Bringmann} \emph {et~al.}}]{Bringmann:2012vr}%
  \BibitemOpen
  \bibfield  {author} {\bibinfo {author} {\bibfnamefont {T.}~\bibnamefont
  {Bringmann}} \emph {et~al.},\ }\href {\doibase 10.1088/1475-7516/2012/07/054}
  {\bibfield  {journal} {\bibinfo  {journal} {JCAP}\ }\textbf {\bibinfo
  {volume} {1207}},\ \bibinfo {pages} {054} (\bibinfo {year} {2012})},\ \Eprint
  {http://arxiv.org/abs/1203.1312} {arXiv:1203.1312 [hep-ph]} \BibitemShut
  {NoStop}%
\bibitem [{\citenamefont {Weniger}(2012)}]{REF:Weniger:2012tx}%
  \BibitemOpen
  \bibfield  {author} {\bibinfo {author} {\bibfnamefont {C.}~\bibnamefont
  {Weniger}},\ }\href {\doibase 10.1088/1475-7516/2012/08/007} {\bibfield
  {journal} {\bibinfo  {journal} {JCAP}\ }\textbf {\bibinfo {volume} {1208}},\
  \bibinfo {pages} {007} (\bibinfo {year} {2012})},\ \Eprint
  {http://arxiv.org/abs/1204.2797} {arXiv:1204.2797 [hep-ph]} \BibitemShut
  {NoStop}%
\bibitem [{\citenamefont {Albert}\ \emph {et~al.}(2014)\citenamefont {Albert}
  \emph {et~al.}}]{REF:LELine}%
  \BibitemOpen
  \bibfield  {author} {\bibinfo {author} {\bibfnamefont {A.}~\bibnamefont
  {Albert}} \emph {et~al.} (\bibinfo {collaboration} {\fermi-LAT
  Collaboration}),\ }\href {\doibase 10.1088/1475-7516/2014/10/023} {\bibfield
  {journal} {\bibinfo  {journal} {JCAP}\ }\textbf {\bibinfo {volume} {1410}},\
  \bibinfo {pages} {023} (\bibinfo {year} {2014})},\ \Eprint
  {http://arxiv.org/abs/1406.3430} {arXiv:1406.3430 [astro-ph.HE]} \BibitemShut
  {NoStop}%
\bibitem [{\citenamefont {{Ackermann}}\ \emph
  {et~al.}(2012{\natexlab{b}})\citenamefont {{Ackermann}} \emph
  {et~al.}}]{REF:2012.P7Perf}%
  \BibitemOpen
  \bibfield  {author} {\bibinfo {author} {\bibfnamefont {M.}~\bibnamefont
  {{Ackermann}}} \emph {et~al.} (\bibinfo {collaboration} {\fermi-LAT
  Collaboration}),\ }\href {\doibase 10.1088/0067-0049/203/1/4} {\bibfield
  {journal} {\bibinfo  {journal} {\apjs}\ }\textbf {\bibinfo {volume} {203}},\
  \bibinfo {eid} {4} (\bibinfo {year} {2012}{\natexlab{b}})},\ \Eprint
  {http://arxiv.org/abs/1206.1896} {arXiv:1206.1896 [astro-ph.IM]} \BibitemShut
  {NoStop}%
\bibitem [{\citenamefont {{Atwood}}\ \emph {et~al.}(2013)\citenamefont
  {{Atwood}} \emph {et~al.}}]{Atwood:2013rka}%
  \BibitemOpen
  \bibfield  {author} {\bibinfo {author} {\bibfnamefont {W.~B.}\ \bibnamefont
  {{Atwood}}} \emph {et~al.} (\bibinfo {collaboration} {\fermi-LAT
  Collaboration}),\ }\href@noop {} {\  (\bibinfo {year} {2013})},\ \Eprint
  {http://arxiv.org/abs/1303.3514} {arXiv:1303.3514 [astro-ph.IM]} \BibitemShut
  {NoStop}%
\bibitem [{\citenamefont {{Navarro}}\ \emph {et~al.}(1996)\citenamefont
  {{Navarro}}, \citenamefont {{Frenk}},\ and\ \citenamefont
  {{White}}}]{REF:1996ApJ...462..563N}%
  \BibitemOpen
  \bibfield  {author} {\bibinfo {author} {\bibfnamefont {J.~F.}\ \bibnamefont
  {{Navarro}}}, \bibinfo {author} {\bibfnamefont {C.~S.}\ \bibnamefont
  {{Frenk}}}, \ and\ \bibinfo {author} {\bibfnamefont {S.~D.~M.}\ \bibnamefont
  {{White}}},\ }\href {\doibase 10.1086/177173} {\bibfield  {journal} {\bibinfo
   {journal} {\apj}\ }\textbf {\bibinfo {volume} {462}},\ \bibinfo {pages}
  {563} (\bibinfo {year} {1996})},\ \Eprint
  {http://arxiv.org/abs/arXiv:astro-ph/9508025} {arXiv:astro-ph/9508025}
  \BibitemShut {NoStop}%
\bibitem [{\citenamefont {{Navarro}}\ \emph {et~al.}(2010)\citenamefont
  {{Navarro}} \emph {et~al.}}]{REF:2010MNRAS.402...21N}%
  \BibitemOpen
  \bibfield  {author} {\bibinfo {author} {\bibfnamefont {J.~F.}\ \bibnamefont
  {{Navarro}}} \emph {et~al.},\ }\href {\doibase
  10.1111/j.1365-2966.2009.15878.x} {\bibfield  {journal} {\bibinfo  {journal}
  {\mnras}\ }\textbf {\bibinfo {volume} {402}},\ \bibinfo {pages} {21}
  (\bibinfo {year} {2010})},\ \Eprint {http://arxiv.org/abs/0810.1522}
  {arXiv:0810.1522} \BibitemShut {NoStop}%
\bibitem [{\citenamefont {{Bahcall}}\ and\ \citenamefont
  {{Soneira}}(1980)}]{REF:1980ApJS...44...73B}%
  \BibitemOpen
  \bibfield  {author} {\bibinfo {author} {\bibfnamefont {J.~N.}\ \bibnamefont
  {{Bahcall}}}\ and\ \bibinfo {author} {\bibfnamefont {R.~M.}\ \bibnamefont
  {{Soneira}}},\ }\href {\doibase 10.1086/190685} {\bibfield  {journal}
  {\bibinfo  {journal} {\apjs}\ }\textbf {\bibinfo {volume} {44}},\ \bibinfo
  {pages} {73} (\bibinfo {year} {1980})}\BibitemShut {NoStop}%
\bibitem [{\citenamefont {Kravtsov}\ \emph {et~al.}(1998)\citenamefont
  {Kravtsov}, \citenamefont {Klypin}, \citenamefont {Bullock},\ and\
  \citenamefont {Primack}}]{Kravtsov:1997dp}%
  \BibitemOpen
  \bibfield  {author} {\bibinfo {author} {\bibfnamefont {A.~V.}\ \bibnamefont
  {Kravtsov}}, \bibinfo {author} {\bibfnamefont {A.~A.}\ \bibnamefont
  {Klypin}}, \bibinfo {author} {\bibfnamefont {J.~S.}\ \bibnamefont {Bullock}},
  \ and\ \bibinfo {author} {\bibfnamefont {J.~R.}\ \bibnamefont {Primack}},\
  }\href {\doibase 10.1086/305884} {\bibfield  {journal} {\bibinfo  {journal}
  {Astrophys.J.}\ }\textbf {\bibinfo {volume} {502}},\ \bibinfo {pages} {48}
  (\bibinfo {year} {1998})},\ \Eprint {http://arxiv.org/abs/astro-ph/9708176}
  {arXiv:astro-ph/9708176 [astro-ph]} \BibitemShut {NoStop}%
\bibitem [{\citenamefont {{Catena}}\ and\ \citenamefont
  {{Ullio}}(2010)}]{REF:2010JCAP...08..004C}%
  \BibitemOpen
  \bibfield  {author} {\bibinfo {author} {\bibfnamefont {R.}~\bibnamefont
  {{Catena}}}\ and\ \bibinfo {author} {\bibfnamefont {P.}~\bibnamefont
  {{Ullio}}},\ }\href {\doibase 10.1088/1475-7516/2010/08/004} {\bibfield
  {journal} {\bibinfo  {journal} {\jcap}\ }\textbf {\bibinfo {volume} {8}},\
  \bibinfo {eid} {004} (\bibinfo {year} {2010})},\ \Eprint
  {http://arxiv.org/abs/0907.0018} {arXiv:0907.0018 [astro-ph.CO]} \BibitemShut
  {NoStop}%
\bibitem [{\citenamefont {Salucci}\ \emph {et~al.}(2010)\citenamefont
  {Salucci}, \citenamefont {Nesti}, \citenamefont {Gentile},\ and\
  \citenamefont {Martins}}]{Salucci:2010qr}%
  \BibitemOpen
  \bibfield  {author} {\bibinfo {author} {\bibfnamefont {P.}~\bibnamefont
  {Salucci}}, \bibinfo {author} {\bibfnamefont {F.}~\bibnamefont {Nesti}},
  \bibinfo {author} {\bibfnamefont {G.}~\bibnamefont {Gentile}}, \ and\
  \bibinfo {author} {\bibfnamefont {C.}~\bibnamefont {Martins}},\ }\href
  {\doibase 10.1051/0004-6361/201014385} {\bibfield  {journal} {\bibinfo
  {journal} {Astron.Astrophys.}\ }\textbf {\bibinfo {volume} {523}},\ \bibinfo
  {pages} {A83} (\bibinfo {year} {2010})},\ \Eprint
  {http://arxiv.org/abs/1003.3101} {arXiv:1003.3101 [astro-ph.GA]} \BibitemShut
  {NoStop}%
\bibitem [{\citenamefont {{Garbari}}\ \emph {et~al.}(2012)\citenamefont
  {{Garbari}}, \citenamefont {{Liu}}, \citenamefont {{Read}},\ and\
  \citenamefont {{Lake}}}]{2012MNRAS.425.1445G}%
  \BibitemOpen
  \bibfield  {author} {\bibinfo {author} {\bibfnamefont {S.}~\bibnamefont
  {{Garbari}}}, \bibinfo {author} {\bibfnamefont {C.}~\bibnamefont {{Liu}}},
  \bibinfo {author} {\bibfnamefont {J.~I.}\ \bibnamefont {{Read}}}, \ and\
  \bibinfo {author} {\bibfnamefont {G.}~\bibnamefont {{Lake}}},\ }\href
  {\doibase 10.1111/j.1365-2966.2012.21608.x} {\bibfield  {journal} {\bibinfo
  {journal} {\mnras}\ }\textbf {\bibinfo {volume} {425}},\ \bibinfo {pages}
  {1445} (\bibinfo {year} {2012})},\ \Eprint {http://arxiv.org/abs/1206.0015}
  {arXiv:1206.0015 [astro-ph.GA]} \BibitemShut {NoStop}%
\bibitem [{\citenamefont {Cowan}\ \emph {et~al.}(2011)\citenamefont {Cowan},
  \citenamefont {Cranmer}, \citenamefont {Gross},\ and\ \citenamefont
  {Vitells}}]{Cowan:2010js}%
  \BibitemOpen
  \bibfield  {author} {\bibinfo {author} {\bibfnamefont {G.}~\bibnamefont
  {Cowan}}, \bibinfo {author} {\bibfnamefont {K.}~\bibnamefont {Cranmer}},
  \bibinfo {author} {\bibfnamefont {E.}~\bibnamefont {Gross}}, \ and\ \bibinfo
  {author} {\bibfnamefont {O.}~\bibnamefont {Vitells}},\ }\href {\doibase
  10.1140/epjc/s10052-011-1554-0, 10.1140/epjc/s10052-013-2501-z} {\bibfield
  {journal} {\bibinfo  {journal} {Eur.Phys.J.}\ }\textbf {\bibinfo {volume}
  {C71}},\ \bibinfo {pages} {1554} (\bibinfo {year} {2011})},\ \Eprint
  {http://arxiv.org/abs/1007.1727} {arXiv:1007.1727 [physics.data-an]}
  \BibitemShut {NoStop}%
\bibitem [{\citenamefont {Buckley}\ \emph {et~al.}(2015)\citenamefont
  {Buckley}, \citenamefont {Charles}, \citenamefont {Gaskins}, \citenamefont
  {Brooks}, \citenamefont {Drlica-Wagner} \emph {et~al.}}]{Buckley:2015doa}%
  \BibitemOpen
  \bibfield  {author} {\bibinfo {author} {\bibfnamefont {M.~R.}\ \bibnamefont
  {Buckley}}, \bibinfo {author} {\bibfnamefont {E.}~\bibnamefont {Charles}},
  \bibinfo {author} {\bibfnamefont {J.~M.}\ \bibnamefont {Gaskins}}, \bibinfo
  {author} {\bibfnamefont {A.~M.}\ \bibnamefont {Brooks}}, \bibinfo {author}
  {\bibfnamefont {A.}~\bibnamefont {Drlica-Wagner}},  \emph {et~al.},\
  }\href@noop {} {\bibfield  {journal} {\bibinfo  {journal} {Phys.Rev.D}\ }
  (\bibinfo {year} {2015})},\ \Eprint {http://arxiv.org/abs/1502.01020}
  {arXiv:1502.01020 [astro-ph.HE]} \BibitemShut {NoStop}%
\bibitem [{\citenamefont {Chernoff}(1938)}]{REF:Chernoff}%
  \BibitemOpen
  \bibfield  {author} {\bibinfo {author} {\bibfnamefont {H.}~\bibnamefont
  {Chernoff}},\ }\href {\doibase doi:10.1214/aoms/1177728725} {\bibfield
  {journal} {\bibinfo  {journal} {Ann. Math. Statist.}\ }\textbf {\bibinfo
  {volume} {25}},\ \bibinfo {pages} {573} (\bibinfo {year} {1938})}\BibitemShut
  {NoStop}%
\bibitem [{\citenamefont {Bruel}(2012)}]{Bruel:2012bt}%
  \BibitemOpen
  \bibfield  {author} {\bibinfo {author} {\bibfnamefont {P.}~\bibnamefont
  {Bruel}},\ }\href {\doibase 10.1088/1742-6596/404/1/012033} {\bibfield
  {journal} {\bibinfo  {journal} {J.Phys.Conf.Ser.}\ }\textbf {\bibinfo
  {volume} {404}},\ \bibinfo {pages} {012033} (\bibinfo {year} {2012})},\
  \Eprint {http://arxiv.org/abs/1210.2558} {arXiv:1210.2558 [astro-ph.IM]}
  \BibitemShut {NoStop}%
\bibitem [{\citenamefont {Ackermann}\ \emph {et~al.}(2014)\citenamefont
  {Ackermann} \emph {et~al.}}]{Ackermann:2014ula}%
  \BibitemOpen
  \bibfield  {author} {\bibinfo {author} {\bibfnamefont {M.}~\bibnamefont
  {Ackermann}} \emph {et~al.} (\bibinfo {collaboration} {\fermi-LAT
  Collaboration}),\ }\href {\doibase 10.1103/PhysRevLett.112.151103} {\bibfield
   {journal} {\bibinfo  {journal} {Phys.Rev.Lett.}\ }\textbf {\bibinfo {volume}
  {112}},\ \bibinfo {pages} {151103} (\bibinfo {year} {2014})},\ \Eprint
  {http://arxiv.org/abs/1403.5372} {arXiv:1403.5372 [astro-ph.HE]} \BibitemShut
  {NoStop}%
\bibitem [{\citenamefont {Hobbs}\ \emph {et~al.}(2006)\citenamefont {Hobbs},
  \citenamefont {Edwards},\ and\ \citenamefont {Manchester}}]{REF:2006:TEMPO2}%
  \BibitemOpen
  \bibfield  {author} {\bibinfo {author} {\bibfnamefont {G.}~\bibnamefont
  {Hobbs}}, \bibinfo {author} {\bibfnamefont {R.}~\bibnamefont {Edwards}}, \
  and\ \bibinfo {author} {\bibfnamefont {R.}~\bibnamefont {Manchester}},\
  }\href {\doibase 10.1111/j.1365-2966.2006.10302.x} {\bibfield  {journal}
  {\bibinfo  {journal} {Mon.Not.Roy.Astron.Soc.}\ }\textbf {\bibinfo {volume}
  {369}},\ \bibinfo {pages} {655} (\bibinfo {year} {2006})},\ \Eprint
  {http://arxiv.org/abs/astro-ph/0603381} {arXiv:astro-ph/0603381 [astro-ph]}
  \BibitemShut {NoStop}%
\bibitem [{\citenamefont {Abdo}(2010)}]{REF:2010.VelaII}%
  \BibitemOpen
  \bibfield  {author} {\bibinfo {author} {\bibfnamefont {A.}~\bibnamefont
  {Abdo}} (\bibinfo {collaboration} {\fermi-LAT}),\ }\href {\doibase
  10.1088/0004-637X/713/1/154} {\bibfield  {journal} {\bibinfo  {journal}
  {Astrophys.J.}\ }\textbf {\bibinfo {volume} {713}},\ \bibinfo {pages} {154}
  (\bibinfo {year} {2010})},\ \Eprint {http://arxiv.org/abs/1002.4050}
  {arXiv:1002.4050 [astro-ph.HE]} \BibitemShut {NoStop}%
\end{thebibliography}%

\end{document}